\documentclass[fleqn,usenatbib]{mnras}
\usepackage{newtxtext,newtxmath}
\usepackage[T1]{fontenc}
\usepackage{ae,aecompl}

\usepackage{graphicx,amssymb,amsmath,amsfonts}

\renewcommand{\mp}{m_{\rm p}}

\newcommand{\kev}{\,\rm keV}

\let\AAold\AA
\renewcommand{\AA}{\text{\AAold}}

\newcommand{\cm}{\,{\rm cm}}

\newcommand{\pc}{\,{\rm pc}}
\newcommand{\kpc}{\,{\rm kpc}}
\newcommand{\Mpc}{\,{\rm Mpc}}

\newcommand{\s}{\,{\rm s}}

\newcommand{\yr}{\,{\rm yr}}

\newcommand{\Gyr}{\,{\rm Gyr}}
\newcommand{\erg}{\,\rm erg}

\newcommand{\K}{\,{\rm K}}

\newcommand{\msun}{\,{\rm M_{\odot}}}
\newcommand{\zsun}{\,{\rm Z_{\odot}}}

\newcommand{\kb}{k_{\rm B}}
\newcommand{\tadv}{t_{\rm flow}}
\newcommand{\cs}{c_{\rm s}}
\newcommand{\vc}{v_{\rm c}}

\newcommand{\vtag}{\frac{{\rm d} \ln v}{{\rm d} \ln r}}
\newcommand{\Ttag}{\frac{{\rm d} \ln T}{{\rm d} \ln r}}
\newcommand{\rhotag}{\frac{{\rm d} \ln \rho}{{\rm d} \ln r}}
\newcommand{\vctag}{\frac{{\rm d} \ln \vc}{{\rm d} \ln r}}
\newcommand{\Lambdatag}{\frac{{\rm d} \ln \Lambda}{{\rm d} \ln T}}
\newcommand{\Lambdatagrho}{\frac{{\rm d} \ln \Lambda}{{\rm d} \ln \rho}}

\newcommand{\mach}{\mathcal{M}}
\newcommand{\Bernoulli}{\mathcal{B}}

\newcommand{\cloudy}{{\sc cloudy}}

\newcommand{\trident}{{\sc trident}}

\newcommand{\Novi}{N_{\text{\sc O\,vi}}}
\newcommand{\Novii}{N_{\text{\sc O\,vii}}}
\newcommand{\Noviii}{N_{\text{\sc O\,viii}}}
\newcommand{\tsim}{t}
\newcommand{\Mstar}{M_*}
\newcommand{\Mdot}{{\dot M}}
\newcommand{\nH}{n_{\rm H}}
\newcommand{\kms}{\,\rm km\ s^{-1}}

\newcommand{\Rvir}{R_{\rm vir}}
\newcommand{\Tvir}{T_{\rm vir}}

\newcommand{\Rsonic}{R_{\rm sonic}}

\renewcommand{\d}{{\rm d}}

\newcommand{\tcool}{t_{\rm cool}}
\newcommand{\Rcool}{R_{\rm cool}}
\newcommand{\Rshock}{R_{\rm shock}}

\newcommand{\Rfeedback}{R_{\rm feedback}}
\newcommand{\Rhalf}{R_{1/2}}

\newcommand{\thubble}{t_{\rm H}}

\newcommand{\tff}{t_{\rm ff}}

\newcommand{\tflow}{\tadv}

\newcommand{\Mhalo}{M_{\rm halo}}

\newcommand{\Tc}{T_{\rm c}}
\newcommand{\fb}{f_{\rm b}}
\newcommand{\HST}{{\it HST}}
\newcommand{\Chandra}{{\it Chandra}}
\newcommand{\XMMNewton}{{\it XMM-Newton}}
\newcommand{\sigrho}{\langle\delta\rho/\rho\rangle_{\rm rms}}

\title[Cooling flow solutions for the CGM]{Cooling flow solutions for the circumgalactic medium}
\author[J. Stern et al.]{
Jonathan Stern,$^{1}$\thanks{CIERA Fellow}\thanks{E-mail: jonathan.stern@northwestern.edu},
Drummond Fielding,$^{2,3}$
Claude-Andr{\'e} Faucher-Gigu{\`e}re$^{1}$
\newauthor
and Eliot Quataert$^{3}$
\\
$^{1}$Department of Physics and Astronomy and CIERA, Northwestern University, Evanston, IL, USA\\
$^{2}$Center for Computational Astrophysics, Flatiron Institute, 162 5th Ave, New York, NY 10010, USA\\
$^{3}$Astronomy Department and Theoretical Astrophysics Center, University of California Berkeley, Berkeley, CA 94720, USA
}

\date{Accepted XXX. Received YYY; in original form ZZZ}

\pubyear{2019}

\begin{document}
\label{firstpage}
\pagerange{\pageref{firstpage}--\pageref{lastpage}}
\maketitle

\begin{abstract}
In several models of galaxy formation feedback occurs in cycles or mainly at high redshift. At times and in regions where feedback heating is ineffective, hot gas in the galaxy halo is expected to form a cooling flow, where the gas advects inward on a cooling timescale.
Cooling flow solutions can thus be used as a benchmark for observations and simulations to constrain the timing and extent of feedback heating. 
Using analytic calculations and idealized 3D hydrodynamic simulations, we show that for a given halo mass and cooling function, steady-state cooling flows form a single-parameter family of solutions, while initially hydrostatic gaseous halos converge on one of these solutions within a cooling time. The solution is thus fully determined once either the mass inflow rate $\Mdot$ or the total halo gas mass are known. 
In the Milky Way (MW) halo, a cooling flow with $\Mdot$ equal to the star formation rate predicts a ratio of the cooling time to the free-fall time of $\sim 10$, similar to some feedback-regulated models. This solution also correctly predicts observed \ion{O}{VII} and \ion{O}{VIII} absorption columns, and the gas density profile implied by \ion{O}{VII} and \ion{O}{VIII} emission. These results suggest ongoing heating by feedback may be negligible in the inner MW halo. Extending similar solutions out to the cooling radius however underpredicts observed \ion{O}{VI} columns around the MW and around other low-redshift star-forming galaxies. 
This can be reconciled with the successes of the cooling flow model with either a mechanism which preferentially heats the \ion{O}{VI}-bearing outer halo, or alternatively if \ion{O}{VI} traces cool photoionized gas beyond the accretion shock. 
We also demonstrate that the entropy profiles of some of the most relaxed clusters are reasonably well described by a cooling flow solution. 
\end{abstract} 


\section{Introduction}

Classic cooling flow solutions were derived in the 1980's following X-ray observations of cluster centers, which revealed gas cooling times shorter than the Hubble time. 
In these spherical solutions the loss of entropy via radiation drives an inflow towards the center of the potential well, at a rate where heating by compression roughly balances radiative energy losses so that the gas temperature remains near the virial temperature (see reviews in \citealt{Fabian+84, Sarazin86}). The inflowing gas is expected to be predominantly single phase, since the flow is advected on the same timescale as the timescale on which thermal instabilities grow (\citealt{MathewsBregman78,Malagoli+87,BalbusSoker89,LiBryan12}).  
Cooling flow solutions have however fallen out of favor in the context of clusters, since the predicted mass flow rate based on the observed X-ray emission exceeds the observed star formation rate in the central galaxy by a factor of $10-100$, and the models predict a central spike in emission which is not observed (see reviews by \citealt{McNamaraNulsen07,Fabian12,McDonald+18}).

Testing similar solutions in halos less massive than clusters has been hampered by the lower emission measure of the virial temperature gas, which drops sharply with decreasing halo mass. Studies have hence been mostly limited to comparing observed star formation rates (SFRs) with estimates of cooling mass flow rates $\Mdot$ based on some assumption on the mass of the cooling circumgalactic gas, say that it is equal to the halo baryon budget (e.g. \citealt{WhiteFrenk91}).
In these studies $\Mdot$ typically overpredicts the observed SFRs. 
However, since in a cooling flow $\Mdot \propto M_{\rm gas}^2$, where $M_{\rm gas}$ is the cooling gas mass, if $M_{\rm gas}$ is overestimated then $\Mdot$ will be overpredicted. The value of $M_{\rm gas}$ could potentially be lower than the halo baryon budget especially at low redshift, due to the effects of strong winds at earlier epochs (e.g., \citealt{Muratov+15, Hafen+18}).
Moreover, most of the cooling gas may be ejected from the galaxy as outflows rather than form stars, in which case the SFRs underestimate $\Mdot$.

In recent years deep X-ray observations with \XMMNewton\ and \Chandra\ have detected absorption and emission from \ion{O}{VII} and \ion{O}{VIII} lines, which most likely originate in the hot gaseous halo of the Milky Way (\citealt{HenleyShelton10, HenleyShelton12, Gupta+12, MillerBregman13, Fang+15, Bregman+18}). The installation of the Cosmic Origins Spectrograph (COS) onboard \HST\ also facilitated surveys of high-ionization ions such as \ion{O}{VI} and \ion{Ne}{VIII} in sightlines through halos of external galaxies (\citealt{ChenMulchaey09,Prochaska+11,Tumlinson+11,Johnson+15,Johnson+17,Keeney+18,Chen+18,Burchett+19}). 
Unless the absorbing gas pressure is a factor of $\sim30$ less than expected in a virial-temperature gaseous halo, these UV absorption features must also trace gas at or near the virial temperature (\citealt{McQuinnWerk18, Stern+18, Burchett+19}). Thus, using these observational constraints, we can test cooling flow solutions in halos less massive than clusters without any assumption on the cooling gas mass or the fraction of it which turns into stars.

Cooling flow solutions are also useful as a benchmark for cosmological and idealized hydrodynamic simulations, in order to disentangle the effects of radiative cooling on gaseous halos from the effects of other physical processes, such as galaxy feedback. Previous studies have typically compared gaseous halos in simulations to hydrostatic solutions (e.g. \citealt{McCarthy+10, Fielding+17, Oppenheimer18}). Since in hydrostatic solutions cooling is neglected, it is not straightforward to discern whether differences between the simulation and the idealized solution are due to cooling or due to other physical effects. Cooling flow solutions bypass this limitation by providing the expected physical properties of gaseous halos if only cooling is present. 

Cooling flows are also an integral phase of time-dependent models of feedback such as feedback limit cycles, in which a cooling flow initially develops, fuels feedback which in turn suppresses the inflow, after which the supply of gas needed to maintain feedback stops and a cooling flow redevelops, and so forth (e.g.~\citealt{CiottiOstriker01,PizzolatoSoker05,LiBryan14a,LiBryan14b,Meece+15,Prasad+15,Prasad+17,Soker16,YangReynolds16,Martizzi+19}). Understanding cooling flows and their observable consequences is hence useful to test this class of models, and to constrain the duty-cycle of their cooling flow phase. 

The goal of this paper is to systematically adapt the cooling flow solutions originally derived for cluster-scale halos, to a wide range of halo masses, down to the scale of dwarf galaxy halos. We include in the solutions updated constraints on the distribution of matter and properties of radiative cooling, which were not available when the original cooling flow solutions where derived. 
We then proceed to test these cooling flow solutions against the recent X-ray and UV constraints mentioned above. We focus on redshift $z\sim0$ halos where observations are most constraining, and defer an analysis of halos at higher redshifts to future work. 
In two companion papers we discuss the implications of cooling flow solutions for the transition between `cold mode' accretion (=`cold flows') and `hot mode' accretion (=`cooling flows'), and compare cooling flow solutions with gaseous halos in the FIRE-2 cosmological simulations (\citealt{Hopkins+18}). These papers are referred to below as Paper II and Paper III, respectively.

This paper is organized as follows. 
In section \ref{s:analytic} we find cooling flow solutions to the steady-state flow equations, first by deriving a self-similar solution akin to \cite{Fabian+84}, and then by direct integration of the flow equations. In section \ref{s:sims} we test the validity of these steady-state solutions using idealized 3D hydrodynamic simulations. Section \ref{s:observations} compares the predictions of cooling flow solutions with available observational constraints. We discuss and compare our results to previous work in section \ref{s:discussion} and summarize in section \ref{s:summary}. 

A flat $\Lambda$CDM cosmology with $H_0=68\kms\Mpc^{-1}$, $\Omega_{\rm M}=0.31$, and a cosmic baryon fraction $\fb=0.158$ is assumed throughout (\citealt{Planck16}).

\section{Steady-state equations}\label{s:analytic}

In this section we solve the spherical steady-state equations for cooling flows. We first present definitions and equations (section \ref{s:basic eqs}), and discuss the range of radii where our simplifying assumptions could be applicable (section \ref{s:applicability}). We then find self-similar solutions (section \ref{s:self similar}) and full numerical solutions to the flow equations (section \ref{s:transonic}). In the last subsections \ref{s:L*}--\ref{s:clusters} we discuss the properties of the solutions for specific halo masses.

\subsection{Equations and definitions}\label{s:basic eqs}

\begin{figure*}
 \includegraphics{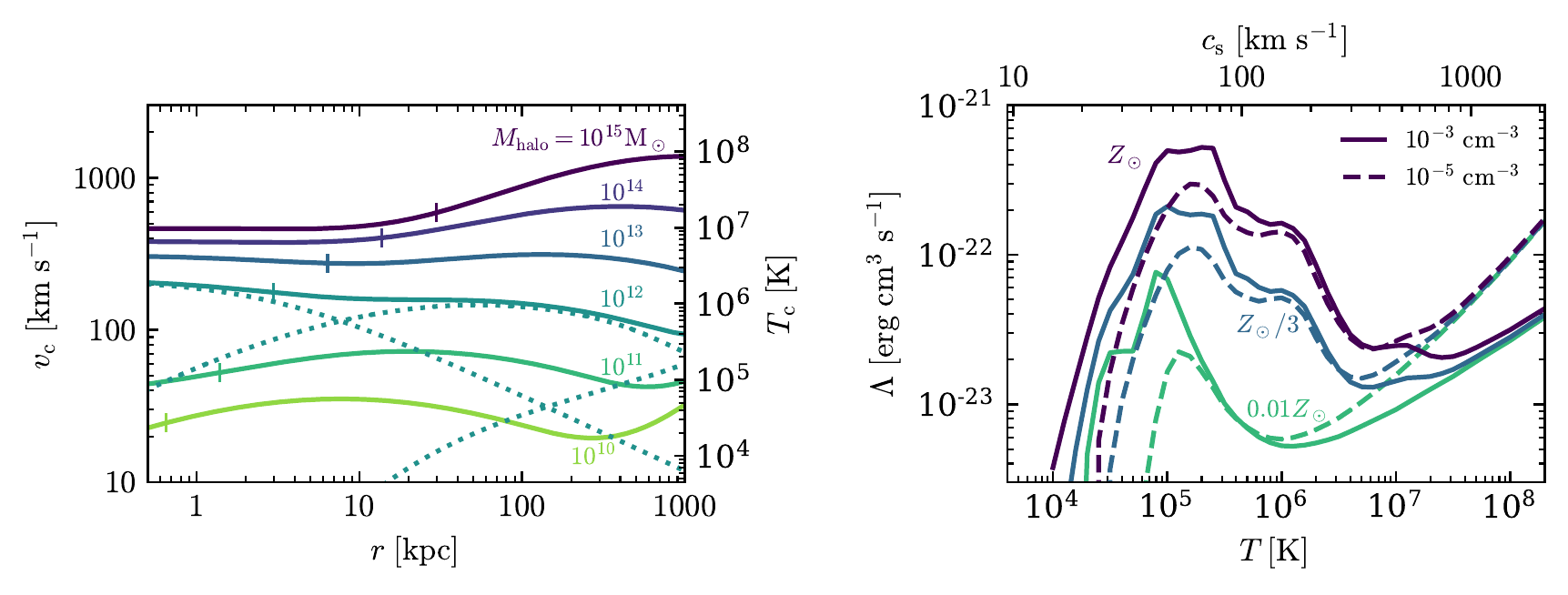}
\caption{\textbf{Left:} Circular velocities ($\vc\equiv\sqrt{GM(<r)/r}$) versus radius in $z=0$ halos. Halo masses are indicated in the panel. We include three mass components for each halo, as plotted by dotted lines for the $10^{12}\msun$ halo: the central galaxy, an NFW halo, and an outer component following Diemer \& Kravtsov (2014). The assumed stellar half-mass radii are noted with small vertical ticks. The right axis is the circular temperature $k\Tc\equiv(3/5)\mu \mp \vc^2$. \textbf{Right:} Cooling functions for gas in $z=0$ dark matter halos, as a function of temperature, metallicity (marked by line color), and density (marked by line style). The adiabatic sound speed is noted on top.}
\label{f:vc and Lambda}
\end{figure*}

The steady-state equations for mass, momentum, and entropy conservation of a spherically-symmetric ideal fluid, with no angular momentum, magnetic field, viscosity, or thermal conduction, are 

\begin{eqnarray}\label{e:mass1}
& \Mdot = 4\pi r^2 \rho v \\
\label{e:momentum1}
& \frac{1}{2}\frac{\d v^2}{\d r} = -\frac{1}{\rho}\frac{\d P}{\d r} - \frac{\vc^2}{r} \\
& v\frac{\d \ln K}{\d r} = -\frac{1}{\tcool}
\label{e:energy1}
\end{eqnarray}
In these equations $r$, $v$, $\rho$, $P$, and $\ln K$  are respectively the radius, radial velocity (negative for an inflow), gas density, gas pressure, and gas entropy ($K\propto P/\rho^\gamma$, where $\gamma$ is the adiabatic index). Also, 
\begin{equation}
 \vc\equiv \sqrt{\frac{GM_{\rm grav}(<r)}{r}}
\end{equation}
is the circular velocity where $M_{\rm grav}(<r)$ is the gravitating mass within $r$, and $\tcool$ is the cooling time, defined as the ratio of the energy per unit volume $(\gamma-1)^{-1}P$ to the radiated energy per unit volume $\nH^2\Lambda$:
\begin{equation}\label{e:tcool}
\tcool = \frac{P}{(\gamma-1)\nH^2 \Lambda} ~.
\end{equation}
where $\nH=X\rho/\mp$ is the hydrogen density ($X$ is the hydrogen mass fraction) and $\Lambda$ is the cooling function. 
We also define the free-fall time
\begin{equation}\label{e:tff}
 \tff = \frac{\sqrt{2}r}{\vc}
\end{equation} and the Bernoulli parameter 
\begin{equation}\label{e:Bernoulli}
 \Bernoulli \equiv \frac{v^2}{2} + \frac{\cs^2}{\gamma-1} + \Phi 
\end{equation}
where $\Phi=-\int (\vc^2/r)\d r$ is the gravitational potential. 

The left panel of Figure~\ref{f:vc and Lambda} plots the values of $\vc(r)$ used in this paper for redshift $z=0$ halos with different masses. The assumed radial mass distributions are composed of three components: an NFW halo, a central galaxy, and an outer component. We ignore the contribution of gas mass to the gravitational mass, validating this assumption {\it a postriori} with our derived gas models. 
For the NFW halo, we calculate the concentration parameter using the relation in \cite{DuttonMaccio14}. The implied virial radius for an overdensity defined as in \cite{BryanNorman98} is
\begin{equation}
 \Rvir = 260 \left(\frac{\Mhalo}{10^{12}\msun}\right)^{1/3} \kpc 
\end{equation}
(we use capital $R$ for radial quantities and small $r$ for the coordinate), while the virial temperature is
\begin{equation}\label{e:Tvir}
 \Tvir= \frac{\mu\mp \vc(\Rvir)^2}{2k} = 6\cdot10^{5}\left(\frac{\Mhalo}{10^{12}\msun}\right)^{2/3} \K  ~.
\end{equation}
The central galaxy is assumed to have a stellar mass $M_*$ estimated from $\Mhalo$ based on the \cite{Behroozi+18} relation for central galaxies. For simplicity, the stellar mass distribution is assumed to have a radial profile:
\begin{equation}
 M_*(<R) = M_*\frac{r}{r+\Rhalf}
\end{equation}
with the stellar half-mass radius $\Rhalf$ taken from \cite{Kravtsov13}:
\begin{equation}\label{e:Rhalf}
 \Rhalf = 0.015 R_{\rm 200c} = 3.0 \left(\frac{\Mhalo}{10^{12}\msun}\right)^{1/3} \kpc.
\end{equation}
Here, $R_{\rm 200c}$ is the radius enclosing an overdensity of 200 relative to the critical density. The values of $\Rhalf$ are marked in the left panel of Fig.~\ref{f:vc and Lambda}. Individual galaxies have a scatter of $0.2\,{\rm dex}$ around this value (\citealt{Kravtsov13}). The value of $\Rhalf$ is also an estimate of the radius where angular momentum may become important (see below). 
For the outer halo component we use the formulation deduced by \citeauthor{DiemerKravtsov14} (2014, hereafter DK14), with their median values of $s_{\rm e}=1.5$ and $b_{\rm e}=1$. 

The right panel of Fig.~\ref{f:vc and Lambda} plots $\Lambda(T, \nH, Z)$ in $z=0$ halos ($T$ and $Z$ are the gas temperature and metallicity, respectively), based on the calculations of \cite{Wiersma+09}, which assume optically thin ionization equilibrium conditions and the UV background of \citeauthor{HaardtMadau12} (2012, hereafter HM12). When calculating $\Lambda$ we use a density which is half the assumed density, since the HM12 background likely underestimates the background by a factor of two (\citealt{FaucherGiguere19}). 

We show below that it is convenient to use the logarithmic derivatives of $\rho$, $v$ and $T$, so we cast the flow equations~(\ref{e:mass1})--(\ref{e:energy1}) in logarithmic form. The mass equation~(\ref{e:mass1}) is equivalent to
\begin{equation}\label{e:dlnrho}
 \frac{\d\ln \rho}{\d\ln r} + \frac{\d\ln v}{\d\ln r} = -2 ~,
\end{equation}
while the entropy equation~(\ref{e:energy1}) can be cast as 
\begin{equation}\label{e:dlnKdlnr}
 \frac{\d\ln T}{\d\ln r} - (\gamma-1)\frac{\d\ln \rho}{\d\ln r} = \frac{\tflow}{\tcool} ~,
\end{equation}
where we used the definition of the flow time 
\begin{equation}\label{e:tflow}
 \tadv\equiv r/|v| ~. 
\end{equation}
Also, defining the adiabatic sound speed 
\begin{equation}\label{e:cs}
 \cs=\sqrt{\frac{\gamma P}{\rho}}
\end{equation}
and multiplying the momentum eqn.~(\ref{e:momentum1}) by $r/\cs^2$ yields:
 \begin{equation}
  \mach^2\frac{\d \ln v}{\d \ln r}  = -\frac{1}{\gamma}\frac{\d\ln P}{\d \ln r} -\frac{\vc^2}{\cs^2}
 \end{equation}
 where $\mach\equiv |v|/\cs$ is the Mach number. Using eqns.~(\ref{e:dlnrho}) and (\ref{e:dlnKdlnr}) to cancel $\d\ln P=\d\ln T+\d \ln \rho$, we get after some rearranging
\begin{equation}\label{e:dlnvdlnr}
  \frac{\d \ln v}{\d \ln r}\left(\mach^2-1\right) = 2 - \frac{\vc^2}{\cs^2} -\frac{\tflow}{\gamma\tcool} ~.
\end{equation}
For $\tcool\rightarrow\infty$ equation~(\ref{e:dlnvdlnr}) reduces to the standard equation used in analyzing adiabatic Bondi flows. 

\subsection{Radii where solutions are applicable}\label{s:applicability}

Before finding solutions to the steady-state equations~(\ref{e:mass1})--(\ref{e:energy1}), we note that the maximum range of radii where we expect them to apply is:
\begin{equation}\label{e:applicability}
 \max{(\Rhalf, \Rfeedback)} \lesssim r \lesssim \min{(\Rcool,\Rshock)}
\end{equation}
where 
$\Rhalf$ is used here as an approximation of the radius where the halo gas could be supported against gravity by angular momentum, 
$\Rfeedback$ is a putative maximum radius where feedback by the galaxy heats, or otherwise alters the physical properties of the circumgalactic gas, 
$\Rcool$ is the usual `cooling radius', where the cooling time equals the age of the system or the time since the last heating event, 
and $\Rshock$ is the radius of the accretion shock. 
We now explain each of these terms.

The $\Rhalf$ limit appears on the left side of eqn.~(\ref{e:applicability}) since the flow equations neglect angular momentum, while centrifugal forces will be significant on the galaxy scale if the specific angular momentum of the gas is similar to the average of the dark matter halo (e.g.~\citealt{Kravtsov13}). 
The feedback radius $\Rfeedback$ is hard to estimate {\it a priori}, and one of the main goals of this study is to provide a benchmark solution in which $\Rfeedback=0$, so deviations from this solution in observations and simulations could be used to constrain $\Rfeedback$. 
The $\Rcool$ limit on the right-hand side of eqn.~(\ref{e:applicability}) is because beyond this radius a steady-state cooling flow does not have time to develop, and a time-dependent solution such as \cite{Bertschinger89} must be found. The value of $\Rcool$ is derived below as a function of halo gas properties.
Last, a discontinuity is expected at $\Rshock$, beyond which the gas is free-falling and close to thermal equilibrium with the UV background, so a cooling flow solution does not apply.

\subsection{Subsonic self-similar solution}\label{s:self similar}

Following \cite{Fabian+84}, we derive a self-similar solution to the flow equations in the subsonic limit $(\mach^2\rightarrow0)$, by approximating $\vc$ as a power-law:
\begin{equation}\label{e:powerlaws}
 \vc(r) = \vc(\Rvir) \left(\frac{r}{\Rvir}\right)^{m}
\end{equation}
where $m$ is constant. For an isothermal potential $m=0$, while around a point mass $m=-0.5$. 
The self-similar solution can then be found by requiring that all logarithmic derivatives of the gas properties are constant, in which case the ratios $\tcool / \tflow$ and $\vc^2 /\cs^2$ are also constant (see eqns.~\ref{e:dlnKdlnr} and \ref{e:dlnvdlnr}). 
For constant $\Lambda$, these two conditions yield (see above definitions of $\tcool$, $\tflow$, and $\cs$):
\begin{equation}\label{e:self similar}
 T\propto r^{2m},~~\nH\propto r^{-\frac{3}{2}+m}, ~~\mach\propto r^{-\frac{1}{2}-2m} 
\end{equation}
where here and henceforth we use $\gamma=5/3$. More general relations can similarly be found for any power-law dependence of $\Lambda$ on $T$, $\nH$, or $r$, though we avoid this complication since it only mildly increases the accuracy of our analytic estimates while significantly increasing the complexity of the analytic expressions. In the numerical integration in the next section we use the full forms of $\vc$ and $\Lambda$ plotted in Fig.~\ref{f:vc and Lambda}. 

Plugging the relations in eqn.~(\ref{e:self similar}) into equations~(\ref{e:dlnKdlnr}) and (\ref{e:dlnvdlnr}) yields
\begin{equation}\label{e:vratio}
 \frac{\vc^2}{\cs^2} = \frac{9}{10}(1-2m) \equiv A~,
\end{equation}
and
\begin{equation}\label{e:tratio}
 \frac{\tflow}{\tcool} = 1 + \frac{4}{3}m \equiv B~.
\end{equation}
Eqn.~(\ref{e:vratio}) implies that the gas temperature in cooling flows is approximately equal to the `circular temperature', defined as
\begin{equation}\label{e:Tc}
 \Tc(r)\equiv \frac{\mu \mp \vc^2(r)}{\gamma \kb}
\end{equation}
which is noted in the left panel of Fig.~\ref{f:vc and Lambda}. We hence get
\begin{equation}\label{e:self similar T}
 T(r) = \frac{1}{A}\Tc(r) = \frac{6}{5A} \left(\frac{r}{\Rvir}\right)^{2m} \Tvir ~.
\end{equation}
where in the second equality we used the definition of $\Tvir$ (eqn.~\ref{e:Tvir}).
Also, eqn.~(\ref{e:tratio}) together with eqn.~(\ref{e:dlnKdlnr}) imply that the entropy profile in cooling flows scales as 
\begin{equation}\label{e:self similar K}
 K\propto r^B\propto r^{1+4m/3} ~.
\end{equation}
This proportionality was previously derived by \cite{McCarthy+05}, and can also be derived from eqn.~(12) in \cite{Voit11}, who found $K\propto r T^{2/3}$ by assuming $\tcool\approx\tflow$ and constant $\Mdot$. Adding the self-similar requirement that $T\propto r^{2m}$ (eqn.~\ref{e:self similar}) then yields eqn.~(\ref{e:self similar K}). 

To complete the solution, we need to derive also the density and Mach number, which depend on the free parameter $\Mdot$. Extracting $\mach$ from $\Mdot = 4\pi r^2\rho v$ (eqn.~\ref{e:mass1}) and $v=r/B\tcool$ (eqn.~\ref{e:tratio}) we get
\begin{equation}\label{e:self similar mach}
 \mach = \frac{XA}{\mp\vc^2}\sqrt{\frac{5\Mdot\Lambda}{18\pi B r}} 
\end{equation}
where we used eqn.~(\ref{e:tcool}) for $\tcool$, eqn.~(\ref{e:vratio}) for $\cs^2$, and we remind the reader that $X$ is the hydrogen mass fraction. Similarly, extracting $\nH$ gives
\begin{equation}\label{e:self similar nH}
\nH = \sqrt{\frac{9B\Mdot}{40\pi A r^3\Lambda}}\vc ~.
\end{equation}  
From equations~(\ref{e:self similar T})--(\ref{e:self similar nH}) and the numerical values of $A$ and $B$ given in eqns.~(\ref{e:vratio})--(\ref{e:tratio}), all the properties of the solution can be found. 
These equations show that for a given $\Mhalo$ and $\Lambda$, cooling flow solutions have a single free parameter. This free-parameter can be chosen according to convenience, and below we use either $\Mdot$, the total gas mass $M_{\rm gas}$, or the sonic radius $\Rsonic$ where $\mach=1$.

An important property of cooling flows is the relation between $\tcool/\tff$ and $\mach$. From eqns.~(\ref{e:vratio})--({\ref{e:tratio}) and the definition of $\tff$ in eqn.~(\ref{e:tff}) we get 
\begin{equation}\label{e:tcool to tff}
 \frac{\tcool}{\tff} = \frac{\tflow/B}{\sqrt{2}r/\vc} = \frac{\sqrt{A}}{\sqrt{2}B}\mach^{-1} ~,
\end{equation}
i.e.\ in cooling flows $\tcool/\tff$ is equal to $\mach^{-1}$ up to a factor of order unity. At $\Rsonic$ we expect $\tcool\approx\tff$.

Below we use these self-similar solutions to derive analytic estimates of the physical properties of cooling flows. We first though compare these solutions to more accurate solutions derived by direct integration.

\begin{figure}
 \includegraphics{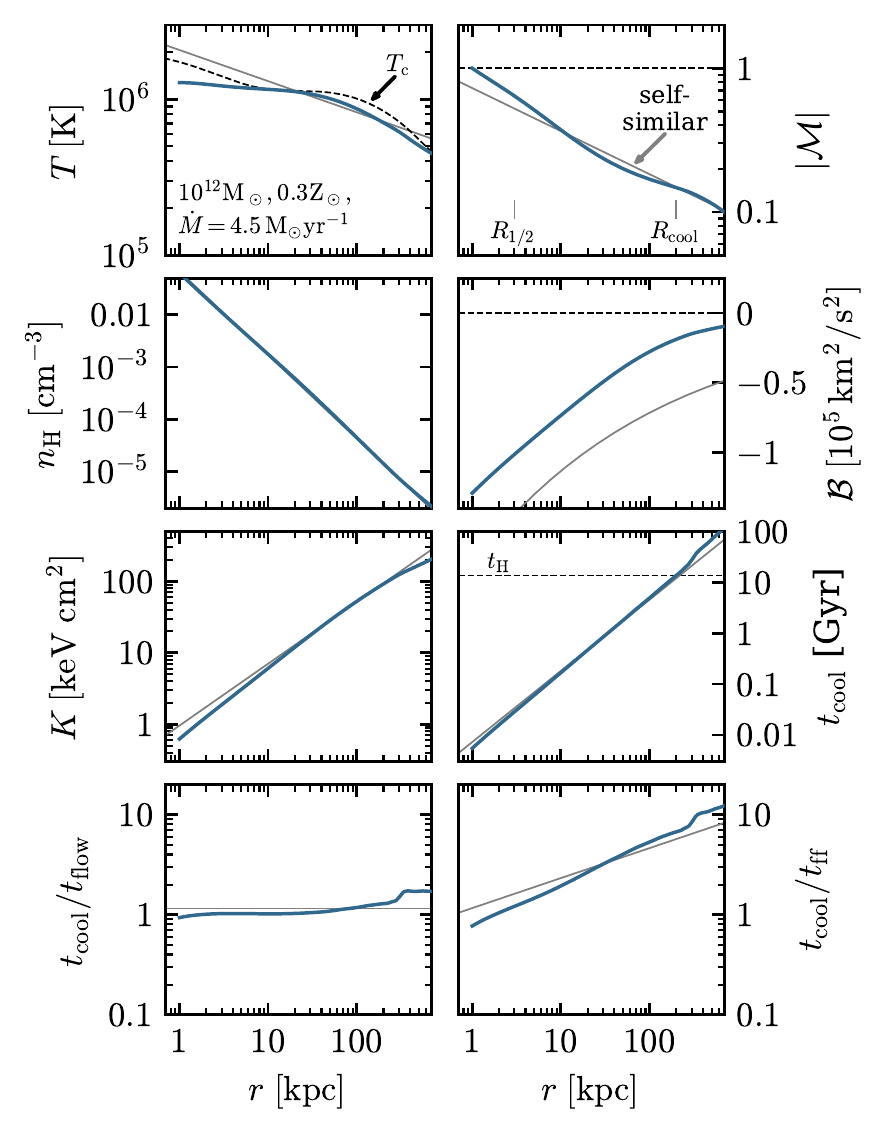}
\caption{An example cooling flow solution for a $10^{12}\msun$ halo at $z=0$. Third-solar metallicity is assumed. Panels show temperature, absolute Mach number, density, Bernoulli parameter, `entropy' ($K\equiv kT/n_{\rm e}$), $\tcool$, $\tcool/\tflow$ where $\tflow\equiv r/v$, and $\tcool/\tff$. 
The blue curves plot a cooling flow solution, derived by integrating the steady-state flow equations~(\ref{e:mass1})--(\ref{e:energy1}) from a sonic point at $\Rsonic=1\kpc$ outward. The solution is required to be marginally bound, i.e.~$\Bernoulli\rightarrow0^{-}$ as $r\rightarrow\infty$, a criterion which yields a single solution per choice of $\Rsonic$. Note that the solution satisfies $T\approx\Tc$ and $\tflow\approx\tcool$. 
The thin gray lines plot an approximate self-similar solution to the flow equations with the same $\Mdot=4.5\msun\yr^{-1}$ as the integrated solution. 
Small marks in the top-right panel denote the stellar half-mass radius $\Rhalf$ and the cooling radius $\Rcool$ where $\tcool=\thubble$. We use $\Rhalf$ as an approximation for the radius where angular momentum may become important in the flow. The radii $\Rhalf$ and $\Rcool$ delimit the maximum range of applicability of the cooling flow solution (see section \ref{s:applicability}). 
}
\label{f:example}
\end{figure}

\begin{figure*}
 \includegraphics{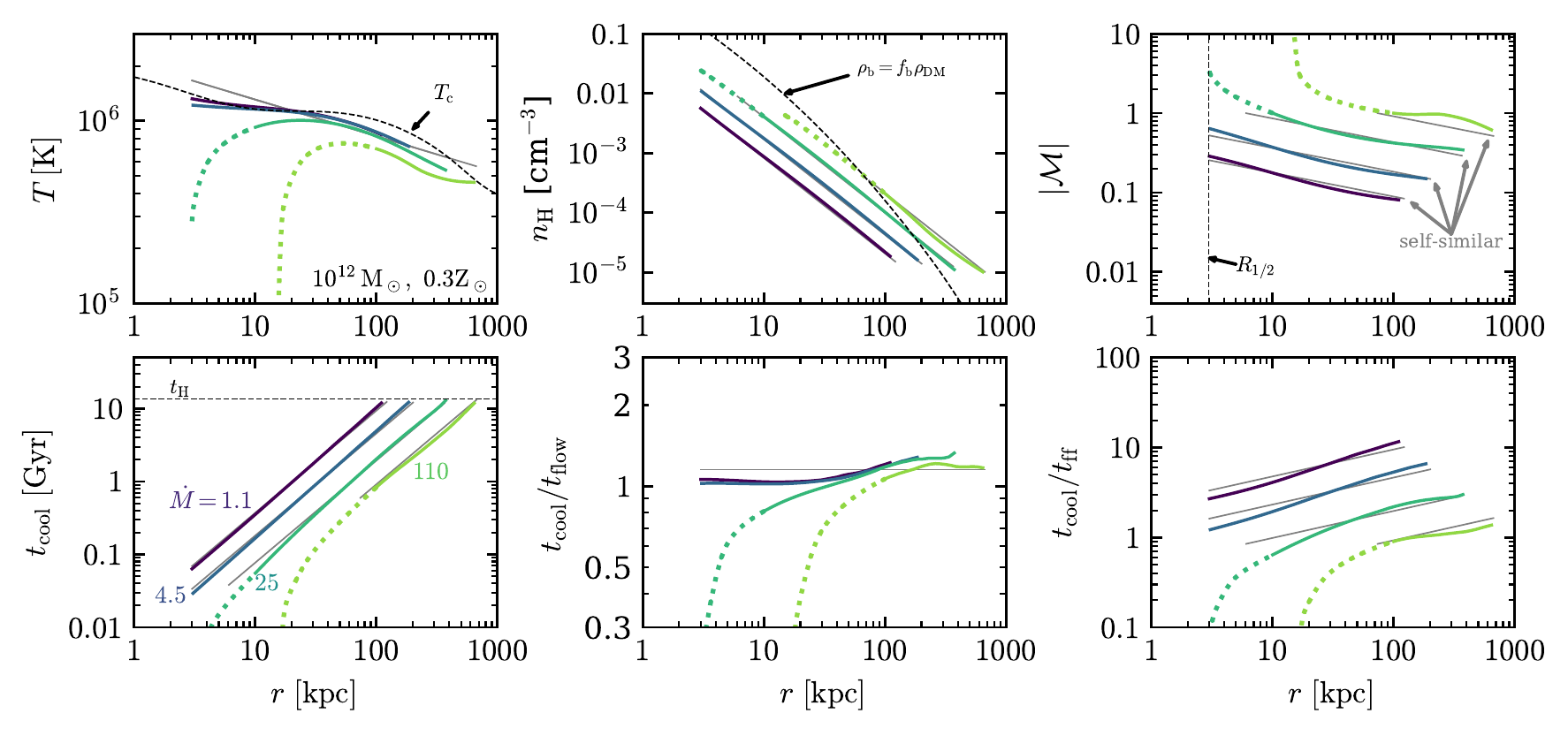}
\caption{Cooling-flow solutions for third-solar metallicity gas in a $10^{12}\msun$ halo at $z=0$. Colored lines plot integrated solutions with $\Rsonic=0.1\kpc$ (purple), $1\kpc$ (blue), $10\kpc$ (green), or $100\kpc$ (light green). The implied $\Mdot$ is noted in $\msun\yr^{-1}$ in the bottom-left panel. Subsonic and supersonic parts of the solutions are denoted by solid  and dotted lines, respectively. 
The curves span the maximum radius range of applicability of the solutions: from $\Rhalf$ (marked in the top-right panel) out to $\Rcool$ ($\thubble$ marked in the lower-left panel). Thin grey lines plot self-similar solutions with the same $\Mdot$ as the integrated solutions. Note the similarity of the self-similar solutions and the subsonic parts of the integrated solutions.
The subsonic parts of the solutions satisfy $T\approx\Tc$ (top-left) and $\tflow\approx\tcool$ (bottom-middle), and have radial slopes of $\nH\propto r^{-1.6}$, $\mach\propto r^{-0.3}$, $\tcool\propto r^{1.4}$ and $\tcool/\tff\propto r^{0.3}$. The normalization of these properties is set by the choice of $\Rsonic$. 
Within $\Rsonic$ the solutions quickly lose their thermal energy and have $\tcool<\tff$ (bottom-right). 
In the top-middle panel the black dashed line marks the gas density if baryons follow the dark matter. 
}
\label{f:by density}
\end{figure*}

\subsection{Integration of flow equations}\label{s:transonic}

Direct integration of the flow equations allows us to account for the effects of a finite $\mach$ on the solution, and also incorporates the full forms of $\vc$ and $\Lambda$ shown in Fig.~\ref{f:vc and Lambda}, rather than the approximations used in the previous section. Similar integrations have previously been done by  \cite{CoxSmith76} and \cite{MathewsBregman78}. 
We start the integration from a sonic point at some radius $\Rsonic$ and integrate outward, allowing $\Rsonic$ to be outside the range of radii of interest (i.e.\ not all solutions are transonic at halo radii). 
At $\Rsonic$ the right side of eqn.~(\ref{e:dlnvdlnr}) must vanish or the velocity derivative will be infinite, hence the three boundary values $\rho(\Rsonic)$, $T(\Rsonic)$ and $v(\Rsonic)$ must satisfy the two conditions $\mach=1$ and $2-\vc^2/\cs^2-\tflow/\gamma\tcool = 0$.  
A third condition on the boundary values is deduced by requiring that the solution is marginally bound, i.e.\ that $\Bernoulli\rightarrow 0^{-}$ as $r\rightarrow\infty$. This choice of the outer boundary condition is motivated by the resemblance to the self-similar solutions discussed in the previous section, which are also marginally-bound.
This boundary condition is enforced on the solution via a shooting method described in Appendix~\ref{a:shooting}, and yields a tight constraint on $T(\Rsonic)$. 
There is thus only a single marginally-bound solution for a given $\Rsonic$, and the entire process yields a one-parameter family of solutions which differ in their value of $\Rsonic$. 

The details of the integrating through the sonic point are also given in Appendix~\ref{a:shooting}. 
We note that other choices of the outer boundary condition, such as an accretion shock or a specific temperature at the cooling radius, change the solution relative to the marginally-bound solution only near the outer boundary, and hence do not affect our conclusions. This insensitivity of the solutions to the exact choice of the boundary condition is demonstrated in Appendix~\ref{a:boundary}. 

In Figure~\ref{f:example} we plot an example transonic marginally-bound solution, for $\Mhalo=10^{12}\msun$, $Z=0.3\zsun$, and an assumed $\Rsonic=1\kpc$. The blue lines in the different panels show different properties of the solution. 
Going inward the temperature roughly follows $\Tc$ (top-left panel), while the entropy drops (third panel on the left). This behavior demonstrates that energy loss to radiation is compensated by adiabatic compression, and hence the effect of radiative cooling is most apparent as a drop in entropy rather than a drop in temperature. 
Gray lines in the panels plot a self-similar solution with the same $\Mdot=4.5\msun\yr^{-1}$ as the integrated solution, using the approximations $\Lambda_{-22} = 0.6$ and $\vc=140 r_{100}^{m}\kms$ with $m=-0.1$, where $r=100\,r_{100}\kpc$ and $\Lambda=10^{-22}\Lambda_{-22}\,{\rm erg}\cm^3\s^{-1}$ (see Fig.~\ref{f:vc and Lambda}). 
With the exception of $\Bernoulli$, the properties of the self-similar and integrated solutions differ by order-unity factors. Specifically, the integrated solution satisfies $\tflow\approx\tcool$ at all radii (lower-left panel), comparable to the constant $\tcool/\tflow=1.15$ in the self-similar solution (eqn.~\ref{e:tratio}). The order unity differences between the integrated and self-similar solution are mainly due to the `decrease-flat-decrease' shape of $\Tc(r)$, which is approximated as a straight power-law in the self-similar solution. The roughly constant offset in $\Bernoulli$ between the two solutions is a result of a constant offset in $\Phi$, since the two solutions differ in the shape of $\Phi(r)$ at large scales.

The radii which limit the range of applicability of the solution, as discussed in section \ref{s:applicability}, are marked in the top-right panel. These radii are $\Rhalf=3\kpc$ (eqn.~\ref{e:Rhalf}), 
and $\Rcool=200\kpc$ derived from the condition $\tcool=\thubble=13.6\Gyr$ ($\thubble$ is marked in the $\tcool$ panel).
We expect the derived solution not to be valid near or outside these limiting radii. 
Near $\Rhalf$ a solution which includes angular-momentum must be found. Around $\Rcool$ one must find a time-dependent solution which accounts for the growth of $\Rcool$ with time. \cite{Bertschinger89} showed that as long as ${\rm d}\Rcool/{\rm d}t \ll \cs(\Rcool)$, such time-dependent solutions join smoothly onto the self-similar solutions. This property is demonstrated in section \ref{s:sims} using hydrodynamic simulations. 
If $\Rshock < \Rcool$ the maximum radius of applicability will be $\Rshock$, and at $\Rshock$ the solution must satisfy the shock jump conditions. In appendix~\ref{a:boundary} we demonstrate that such a solution is essentially identical to the marginally-bound solution within $\Rshock$.

\subsection{Cooling flows in galaxy-scale halos}\label{s:L*}

In this section we discuss several properties of cooling flows in halos characteristic of $\sim$$L^*$ galaxies ($\Mhalo \sim 10^{12}\msun$), using both analytic estimates based on the self-similar solutions (section \ref{s:self similar}) and the more accurate integrated solutions (section \ref{s:transonic}). Figure~\ref{f:by density} plots four integrated and four self-similar solutions for $Z=0.3\zsun$, and $M_{12}=1$, where we define $\Mhalo=10^{12}M_{12}\msun$. For the integrated solutions we assume $\Rsonic=0.1\kpc$ (purple), $1\kpc$ (blue), $10\kpc$ (green), and $100\kpc$ (light green), while for the self-similar solutions (thin grey lines) we use the values of $\Mdot$ found in the integrated solutions, as noted in the bottom-left panel. The blue solution with $\Rsonic=1\kpc$ is the solution shown in Fig.~\ref{f:example}. 
The curves span the maximum range of applicability of the solutions, between $\Rhalf$ and $\Rcool$. 
In cases where $\Rsonic>\Rhalf$, dotted lines plot the inner supersonic part of the solution. 
We first address the subsonic part of the solutions which are the focus of this paper, and then address the supersonic part.

The self-similar solutions provide a good approximation to the subsonic part of the integrated solutions. Specifically, the integrated solutions satisfy $T\approx \Tc$ (top-left panel) and $\tflow\approx\tcool$ (bottom-middle), as expected from eqns.~(\ref{e:vratio})--(\ref{e:tratio}) derived in the context of the self-similar solutions. These two conditions demonstrate why cooling flows form a single-parameter family of solutions. The free parameter sets the normalization of other plotted properties -- $\nH$, $\mach$, $\tcool$, and $\tcool/\tff$. The slope of the profiles of these properties is however roughly the same in all solutions. 

\begin{figure*}
 \includegraphics{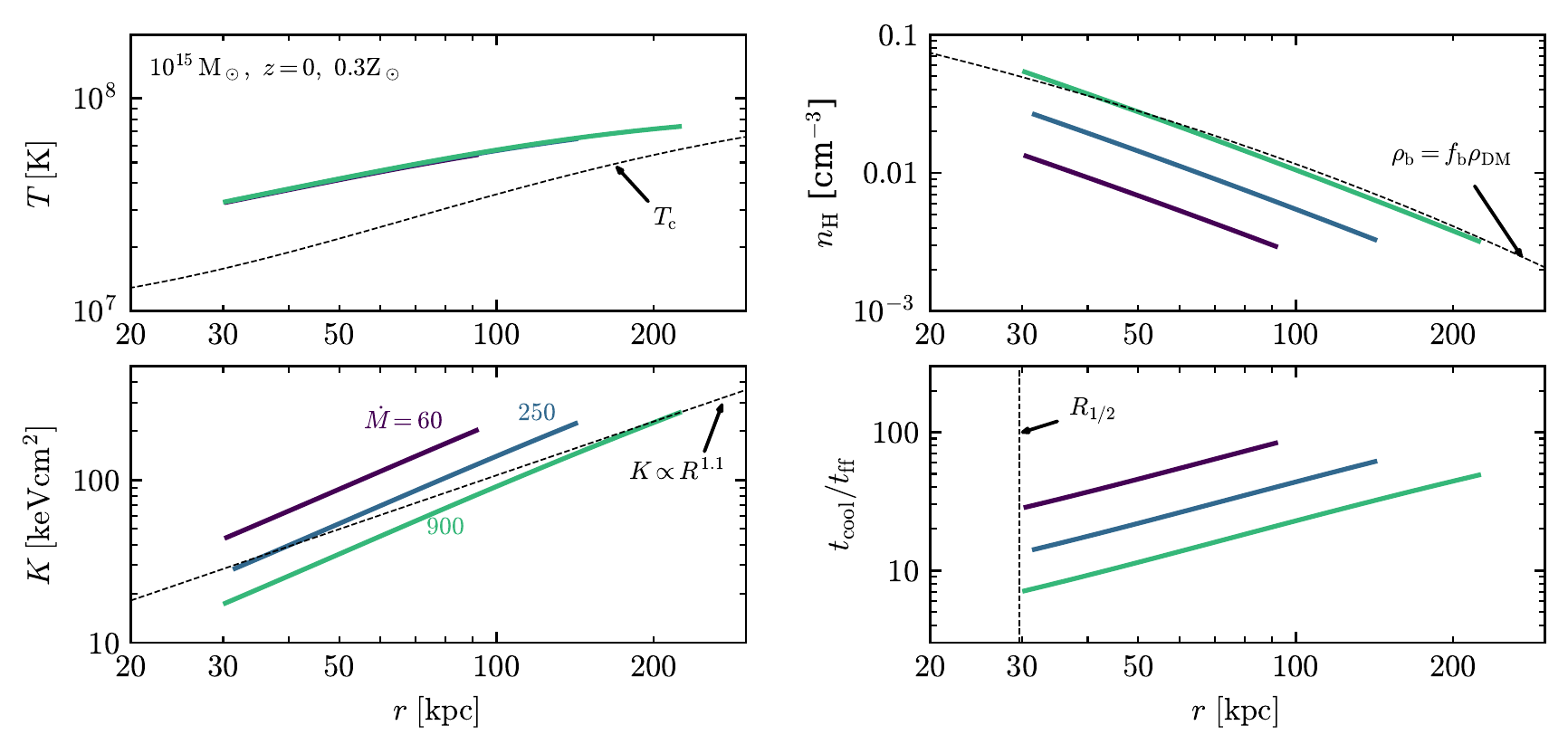}
\caption{
Cooling-flow solutions for third-solar metallicity gas in a $10^{15}\msun$ halo at $z=0$. 
The values of $\Mdot$ are noted in the bottom-left panel in $\msun\yr^{-1}$. 
The curves span from the galaxy scale $\Rhalf=30\kpc$ to the cooling radius $\Rcool$.
\textbf{(Top-left)} Gas temperature. The solutions have a positive $T$ slope, though flatter than $\Tc$ (black dashed line). The value of $\Mdot$ has a negligible affect on $T$. 
\textbf{(Top-right)}
Gas density. The solutions have a similar density slope as the dark matter (black dashed line). For a baryon to dark matter mass ratio equal to the cosmic baryon fraction (green line) the expected $\Mdot$ is $900\msun\yr^{-1}$. 
\textbf{(Bottom-left)}
Entropy. Cooling flows have $K\propto R^{1.4}$ (eqn.~\ref{e:entropy cluster}), steeper than $K\propto R^{1.1}$ expected from self-similar cosmological accretion without cooling (black dashed line). 
\textbf{(Bottom-right)}
Cooling time to free-fall time. Cooling flows in clusters remain stable ($\tcool>\tff$) down to $\Rhalf$.
}
\label{f:M15}
\end{figure*}

The top-middle panel of Fig.~\ref{f:by density} shows the density profile of the solutions. For the self-similar solutions, this profile can also be derived analytically. Approximating $\vc$ as $140\, M_{12}^{0.36}r_{100}^{m}\kms$ with $m=-0.1$ (Fig.~\ref{f:vc and Lambda}), eqns.~(\ref{e:vratio})--(\ref{e:tratio}) imply $A=1.08$ and $B=0.87$, so we get from eqn.~(\ref{e:self similar nH}) that
\begin{equation}\label{e:nH Lstar}
 \nH = 1.6 \cdot 10^{-5}M_{12}^{0.36}\, \Mdot_1^{1/2}\, \Lambda_{-22}^{-1/2}\, r_{100}^{-1.6}  \cm^{-3} ~,
\end{equation}
where $\Mdot=1 \Mdot_1 \msun\yr^{-1}$, and we used $X=0.75$. This normalization of $\Mdot$ is based on the SFR of the Milky Way (\citealt{BlandHawthornGerhard16}), and corresponds to the purple lines in Fig.~\ref{f:by density}. 
Eqn.~(\ref{e:nH Lstar}) implies that $\Mdot\propto \nH^2$, as expected in cooling flows since $\Mdot\propto \rho v$ and $v\approx r/ \tcool\propto \rho$. 
Also, the scaling $\nH\propto r^{-1.6}$ is consistent with estimates of the hot gas density slope in the Milky Way halo based on X-ray emission and absorption (\citealt{Bregman+18}). We return to this comparison in section \ref{s:observations}. 

The dashed black line in the top-middle panel plots the expected density if baryons follow the dark matter with a cosmic baryon fraction $\fb$. This estimate is higher than the $\Mdot=1\msun\yr^{-1}$ solution by an order of magnitude at $\Rcool$ (the right end of the plotted line), and by a larger factor at smaller radii. Hence the $\Mdot=1\msun\yr^{-1}$ solution corresponds to a highly baryon-depleted halo. 

Using eqns.~(\ref{e:tcool}), (\ref{e:self similar T}) and (\ref{e:nH Lstar}), we can derive an expression for $\tcool$:
\begin{equation}\label{e:tcool Lstar}
 \tcool = 7.2 \, M_{12}^{0.36} \Mdot_1^{-1/2}\, \Lambda_{-22}^{-1/2} r_{100}^{1.4} \Gyr ~,
\end{equation}
which can be compared to the integrated solutions in the lower-left panel of Fig.~\ref{f:by density}. The cooling radius where $\tcool=13.6\Gyr$ is hence
\begin{equation}\label{e:Rcool Lstar}
\Rcool = 130 \, M_{12}^{-0.26} \Mdot_1^{0.36}\, \Lambda_{-22}^{0.36}\kpc ~.
\end{equation}  
Equation~(\ref{e:Rcool Lstar}) provides an estimate of the outer limit of applicability of the cooling flow solution. 

Similarly, the Mach number profile in cooling flows can be derived from eqn.~(\ref{e:self similar mach}): 
\begin{equation}\label{e:mach Lstar}
 \mach = 0.11\, M_{12}^{-0.72}\, \Mdot_1^{1/2}\, \Lambda_{-22}^{1/2}\, r_{100}^{-0.3} ~,\\ 
\end{equation}
and the ratio $\tcool/\tff$ in cooling flows follows from eqn.~(\ref{e:tcool to tff}) 
\begin{equation}\label{e:tcool to tff Lstar}
 \frac{\tcool}{\tff} = 7.5\, M_{12}^{0.72}\, \Mdot_1^{-1/2}\, \Lambda_{-22}^{-1/2}\, r_{100}^{0.3} ~.
\end{equation}
Eqns.~(\ref{e:mach Lstar})--(\ref{e:tcool to tff Lstar}) imply that $\tcool/\tff \propto \mach^{-1}$ is a weak function of radius in $\sim$$L^*$ halos, as can be seen in the right panels of Fig.~\ref{f:by density}. 
The flatness of $\tcool/\tff$ is similar to the basic ansatz of thermal instability and `precipitation' models based on simulations with galaxy feedback (\citealt{Sharma+12b}, \citealt{Voit+17}), though note that the physics are different since the cooling flow solution does not include feedback. 
We compare cooling flow solutions to precipitation models in section~\ref{s:discussion}. 

We now turn to discuss the supersonic parts of the solutions. 
A rough estimate of $\Rsonic$ where $\mach=1$ can be derived from eqn.~(\ref{e:mach Lstar}), which yields:
\begin{equation}\label{e:Rsonic Lstar}
 \Rsonic = 0.06 M_{12}^{-2.4}\, \Mdot_1^{1.67}\, \Lambda_{-22}^{1.67} \kpc~.
\end{equation}
This estimate can be compared to the integrated results in the top-right panel of Fig.~\ref{f:by density}. Within $\Rsonic$, the solutions rapidly lose their thermal energy (top-left panel) and the solution becomes a free-falling solution rather than a cooling flow solution. Within $\Rsonic$ the solution also has $\tcool<\tff$ (bottom-right panel and eqn.~\ref{e:tcool to tff}), which is incompatible with the assumption of steady-state used to derive these solutions. That is, within $\Rsonic$ the flow is likely to be subject to 
thermal instabilities, repressurizing shocks, and inefficient mixing, so we do not expect the steady-state solutions to apply. 

For solutions with $\Rsonic>\Rhalf$ (e.g.\ the green and light green solutions in Fig.~\ref{f:by density}), one may ask whether the subsonic part of the solution at $R>\Rsonic$ is valid, if there is no valid steady-state solution within $\Rsonic$ which can support its weight against gravity? 
In Paper II we show that indeed, if $\Rsonic>\Rhalf$ 
all the halo gas collapses on a dynamical timescale, including the subsonic gas beyond $\Rsonic$. Steady-state solutions for the halo gas are thus possible only if the gas remains subsonic down to $\Rhalf$.

\subsection{Cooling flows in cluster-scale halos}\label{s:clusters}

Figure~\ref{f:M15} plots three cooling flow solutions in $10^{15}\msun$ halos, derived assuming $Z=0.3\zsun$ and either $\Rsonic=0.1\kpc$ (purple), $0.3\kpc$ (turquoise) or $1\kpc$ (green). The implied $\Mdot$ are noted in the bottom-left panel. In cluster-scale halos the density at $\Rcool$ is directly observed via its X-ray emission (see section~\ref{s:observations}), and is found to be roughly consistent with the cosmic baryon fraction, corresponding to the normalization of the green solution in Fig.~\ref{f:M15}. 

To derive the self-similar solution for cluster-scale halos, we approximate $\vc$ as $850\,M_{15}^{0.23} r_{100}^m \kms$ with $m\approx0.3$, where $\Mhalo=10^{15}M_{15}\msun$ (see Fig.~\ref{f:vc and Lambda}). In this case eqns.~(\ref{e:vratio})--(\ref{e:tratio}) imply $A=0.36$ and $B=1.4$. Hence in the self-similar solution we get (eqn.~\ref{e:self similar T})
\begin{equation}\label{e:T cluster}
 T = 2.8 \Tc
\end{equation}
The top-left panel in Fig.~\ref{f:M15} shows that in the integrated solution $T$ is a factor of $1.5-2$ above $\Tc$, less than suggested by eqn.~(\ref{e:T cluster}),  and has a weaker dependence on radius than $\Tc$. These differences between the integrated and self-similar solutions occur because of the flattening of the $\vc$ profile beyond $300\kpc$ (see Fig.~\ref{f:vc and Lambda}), which affects the integrated solution but is not captured by the power-law approximation of the gravitational potential in the self-similar solution. 

Calculating eqn.~(\ref{e:self similar nH}) in the context of clusters gives
\begin{equation}\label{e:nH cluster}
 \nH = 2.1 \cdot 10^{-2}M_{15}^{0.23}\, \Mdot_{1000}^{1/2}\, \Lambda_{-23}^{-1/2}\, r_{100}^{-1.2}  \cm^{-3}
\end{equation}
where we defined $\Lambda=10^{-23}\Lambda_{-23}\erg\cm^3\s^{-1}$ and $\Mdot=10^3 \Mdot_{1000}\msun\yr^{-1}$. For comparison, the integrated solutions suggest a somewhat steeper slope of $\nH\propto r^{-1.4}$ (top-right panel of Fig.~\ref{f:M15}). This difference is also due to the flattening of $\vc$ beyond $300\kpc$. Note that the density slope in cooling flows is similar to the dark matter slope at these radii. The cooling time is
\begin{equation}\label{e:tcool cluster}
 \tcool = 6.1 \, M_{15}^{0.23} \Mdot_{1000}^{-1/2}\, \Lambda_{-23}^{-1/2} r_{100}^{1.8} \Gyr ~,
\end{equation}
while the cooling radius is
\begin{equation}\label{e:Rcool cluster}
\Rcool = 160 \, M_{15}^{-0.13} \Mdot_{1000}^{0.28}\, \Lambda_{-23}^{-0.28}\kpc ~.
\end{equation}
The value of $\Rcool$ is similar to that in $\sim$$L^*$ galaxies (eqn.~\ref{e:tcool Lstar}).

\begin{figure*}
   \centering
    \begin{minipage}{0.24\textwidth}
        \centering
        \includegraphics[width=\textwidth]{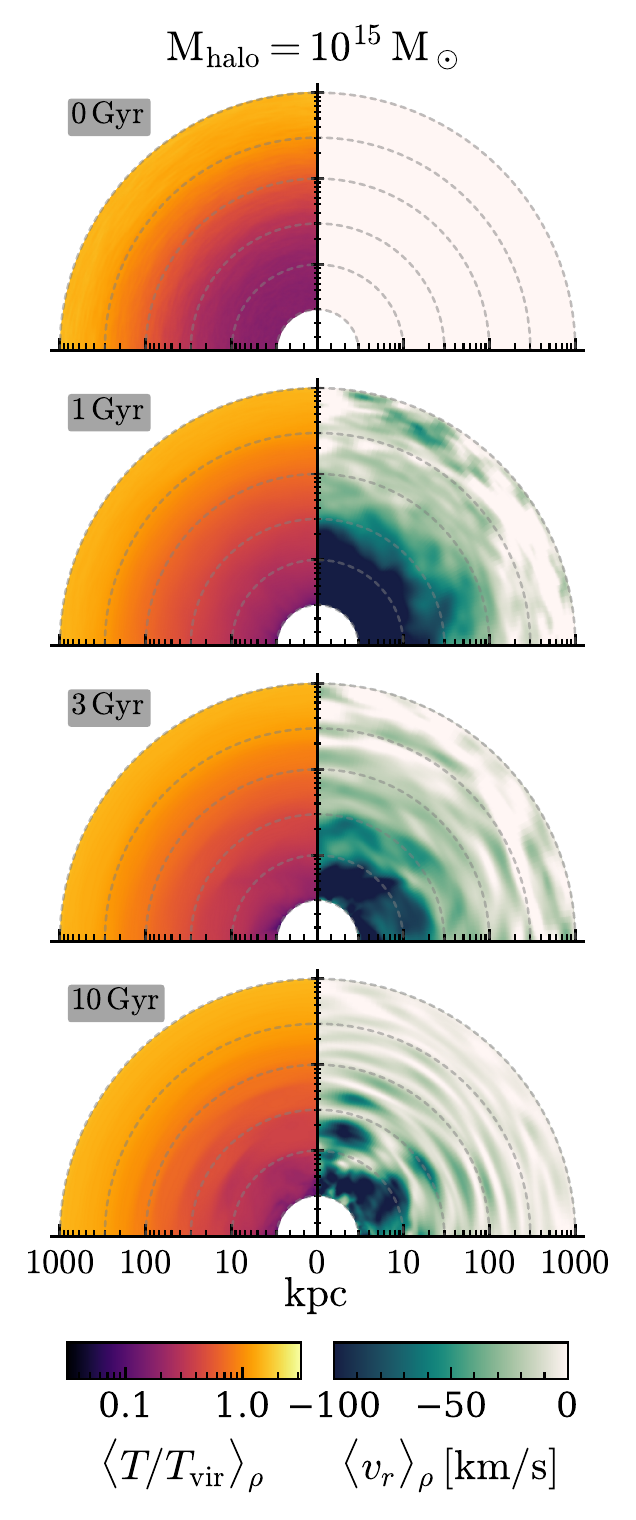}
    \end{minipage}
    \begin{minipage}{0.24\textwidth}
        \centering
        \includegraphics[width=\textwidth]{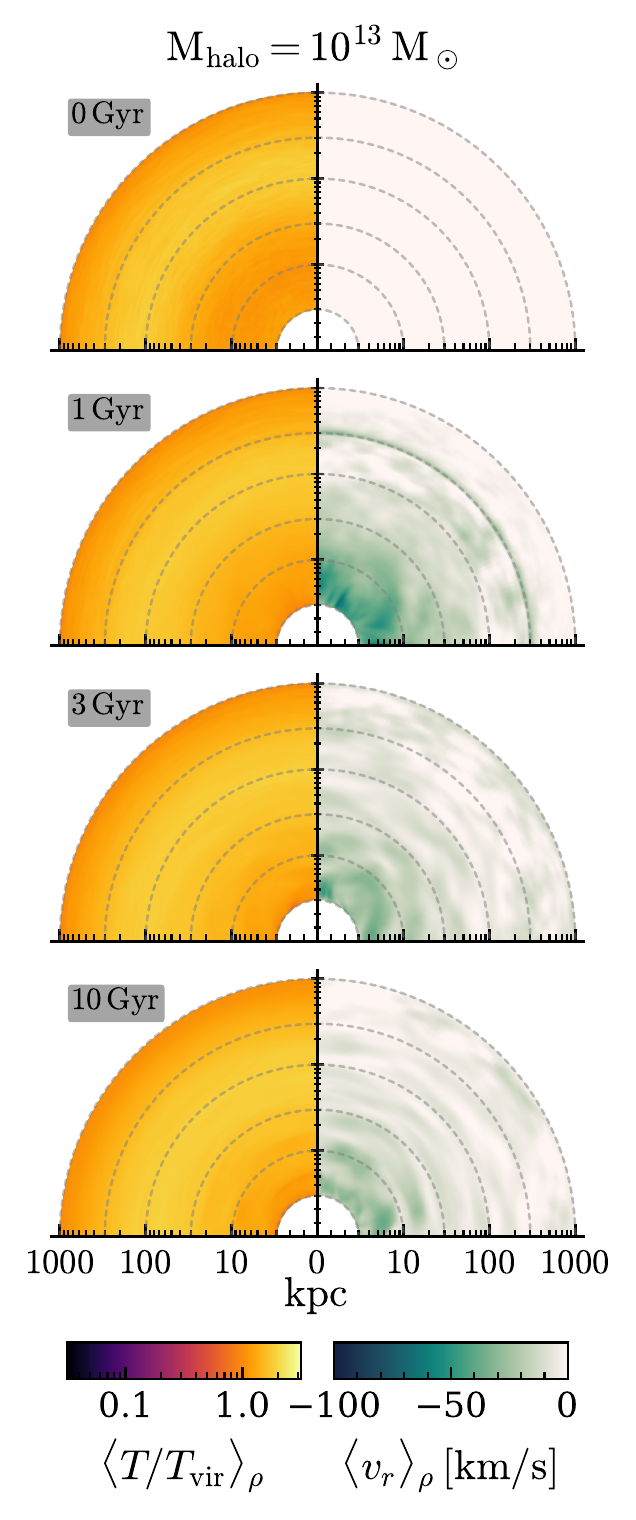}
    \end{minipage}
    \begin{minipage}{0.24\textwidth}
        \centering
        \includegraphics[width=\textwidth]{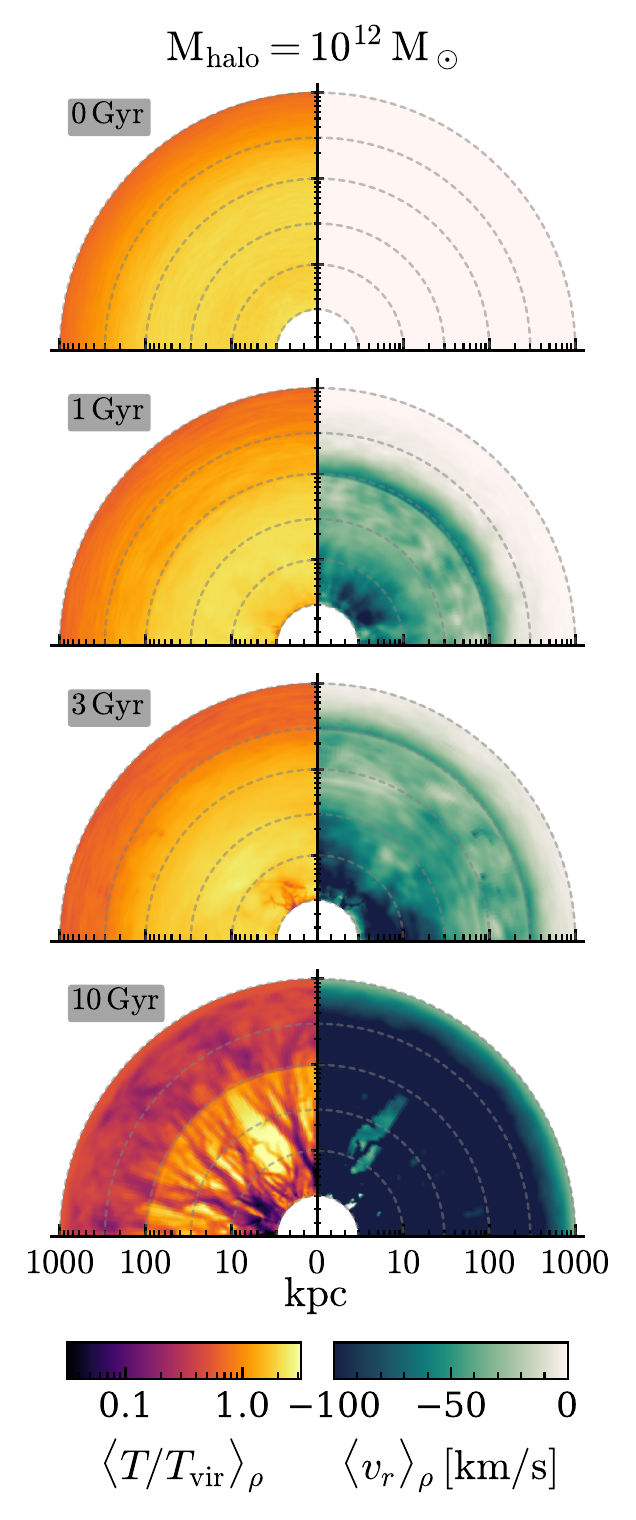}
    \end{minipage}
    \begin{minipage}{0.24\textwidth}
        \centering
        \includegraphics[width=\textwidth]{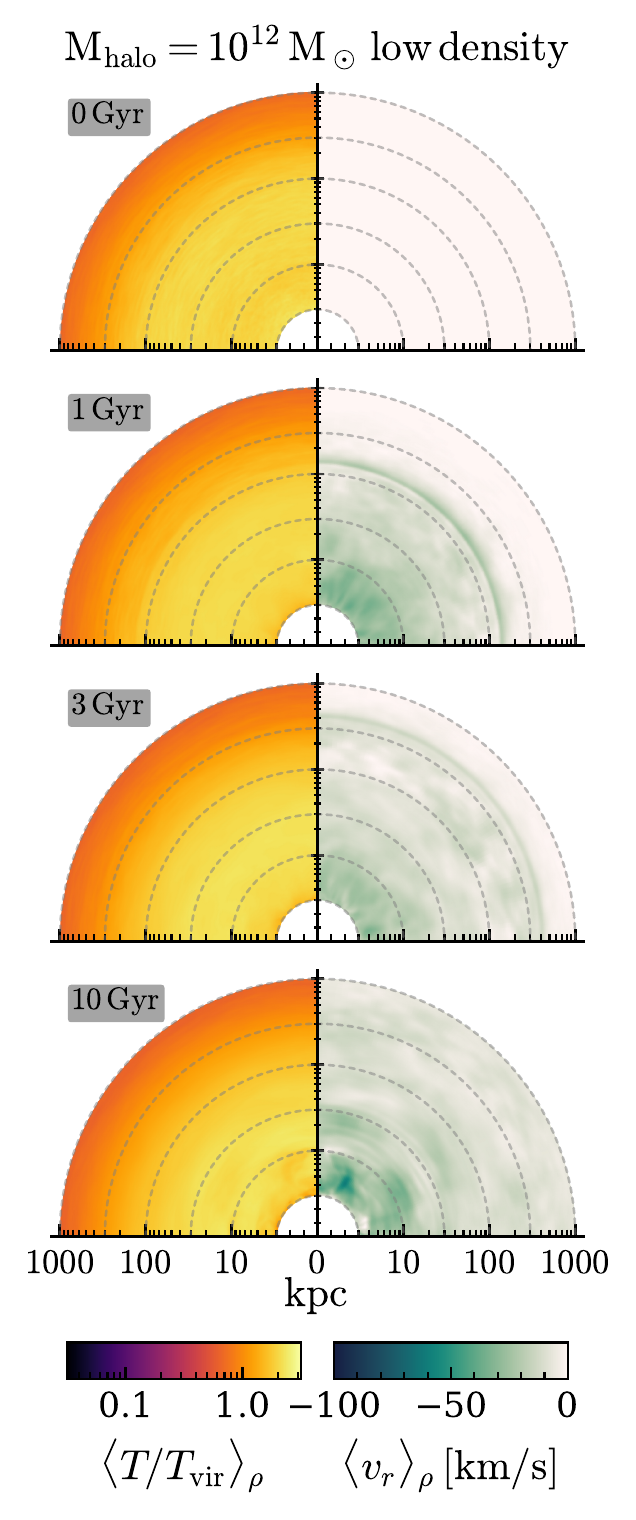}
    \end{minipage}
\caption{Idealized 3D hydro simulations of radiatively cooling gas in a dark matter halo. Halo mass is noted on top, and $Z=\zsun/3$ is assumed throughout. The simulations in the three left columns assume an initial gas mass equal to $\fb\Mhalo-\Mstar$, while the simulation in the right column has an initial gas mass equal to a third of this value. In each column we plot snapshots of temperature (in units of $\Tvir$) and radial velocity at different simulation times. The snapshots are plotted in the $r-\theta$ plane, averaged (mass-weighted) over the $\phi$ coordinate. The initial conditions are hydrostatic (top row). Note how an inflow develops in all simulations. Significant deviations from spherical symmetry appear only at late times in the fiducial density $10^{12}\msun$ simulation. 
}
\label{f:snapshots}
\end{figure*}

The entropy of the hot gas can be estimated from eqns.~(\ref{e:T cluster}) and (\ref{e:nH cluster}):
\begin{equation}\label{e:entropy cluster}
 K \equiv \frac{kT}{(n_{\rm e})^{2/3}} =  103 \, M_{15}^{0.31} \Mdot_{1000}^{-1/3}\, \Lambda_{-23}^{1/3}\, r_{100}^{1.4} \kev \cm^2
\end{equation}
where $n_{\rm e}\approx1.2\,\nH$ is the electron density. The derived slope is consistent with the slope found in the integrated solutions, shown in the bottom-left panel of Fig.~\ref{f:M15}. 
Moreover, this slope is steeper than the $K\propto r^{1.1}$ expected from self-similar cosmological accretion theory in the absence of cooling (e.g.\ \citealt{TozziNorman01}). Therefore, in the absence of physical processes other than gravity and cooling, the entropy profile is expected to steepen around $\Rcool$ from $\sim r^{1.1}$ to $ \sim r^{1.4}$. Note that the entropy profile scales as $\sim r^{1+4/3m}$ (eqn.~\ref{e:self similar K}) and $m$ becomes smaller with decreasing mass (Fig.~\ref{f:vc and Lambda}), so the expected steepening at $\Rcool$ is weaker in group-scale halos than in cluster-scale halos. 

Last, we calculate $\tcool/\tff$ in clusters. Using eqn.~(\ref{e:tcool to tff}) we get
\begin{equation}\label{e:tcool to tff clusters}
 \frac{\tcool}{\tff} = 38\, M_{15}^{0.46}\, \Mdot_{1000}^{-1/2}\, \Lambda_{-23}^{-1/2}\, r_{100}^{1.1} ~.\\ 
\end{equation}
This relation can be compared to the bottom-right panel of Fig.~\ref{f:M15}, where we plot $\tcool/\tff$ in the integrated solutions.
Eqn.~(\ref{e:tcool to tff clusters}) implies that cooling flows in clusters are expected to remain stable ($\tcool>\tff$) down to the galaxy scale $\Rhalf\approx30\kpc$.

\section{Hydrodynamic simulations}\label{s:sims}

In this section we present a series of idealized 3D hydrodynamic simulations of gas in dark matter halos. Concisely put, our simulations are controlled numerical experiments in which gas that is initially in hydrostatic equilibrium with an external gravitational potential is allowed to cool radiatively. Roughly hydrostatic conditions are expected in realistic halos after a virial shock has formed (see further discussion in Paper II). Our goal is to verify that in a hydrodynamic setting the halo gas indeed chooses one of the single-parameter family of solutions discussed in the previous section. The simulations are based on the simulations described in \cite{Fielding+17}, though they do not include feedback from the central galaxy. 

\subsection{Simulation setup}

The simulations are performed using the grid-based hydrodynamics code \textsc{athena++} (Stone et al.\ in prep). We adopt an adiabatic equation of state with $\gamma = 5/3$ and solve the standard hydrodynamics equations with additional source terms to include gravity and optically thin cooling and photoionization heating. 
We impose a cooling time constraint on the time step so that the global time step is the smaller of $\delta t_{\rm hydro}$ and ${\rm min}\{\tcool\}/10$, to ensure that cells do not over-cool in one time step. Both the cooling and hydrodynamics updates are done with this time step.
The same \cite{Wiersma+09} cooling and heating tables are used in the simulations as in the analytic calculations in the previous section (examples are shown in right panel of Fig.~\ref{f:vc and Lambda}). Likewise, the simulations use the same gravitational accelerations that give rise to the $\vc$ profiles shown in the left panel of Fig.~\ref{f:vc and Lambda}. 

\subsubsection{Simulation domain and geometry}

\begin{figure*}
 \includegraphics{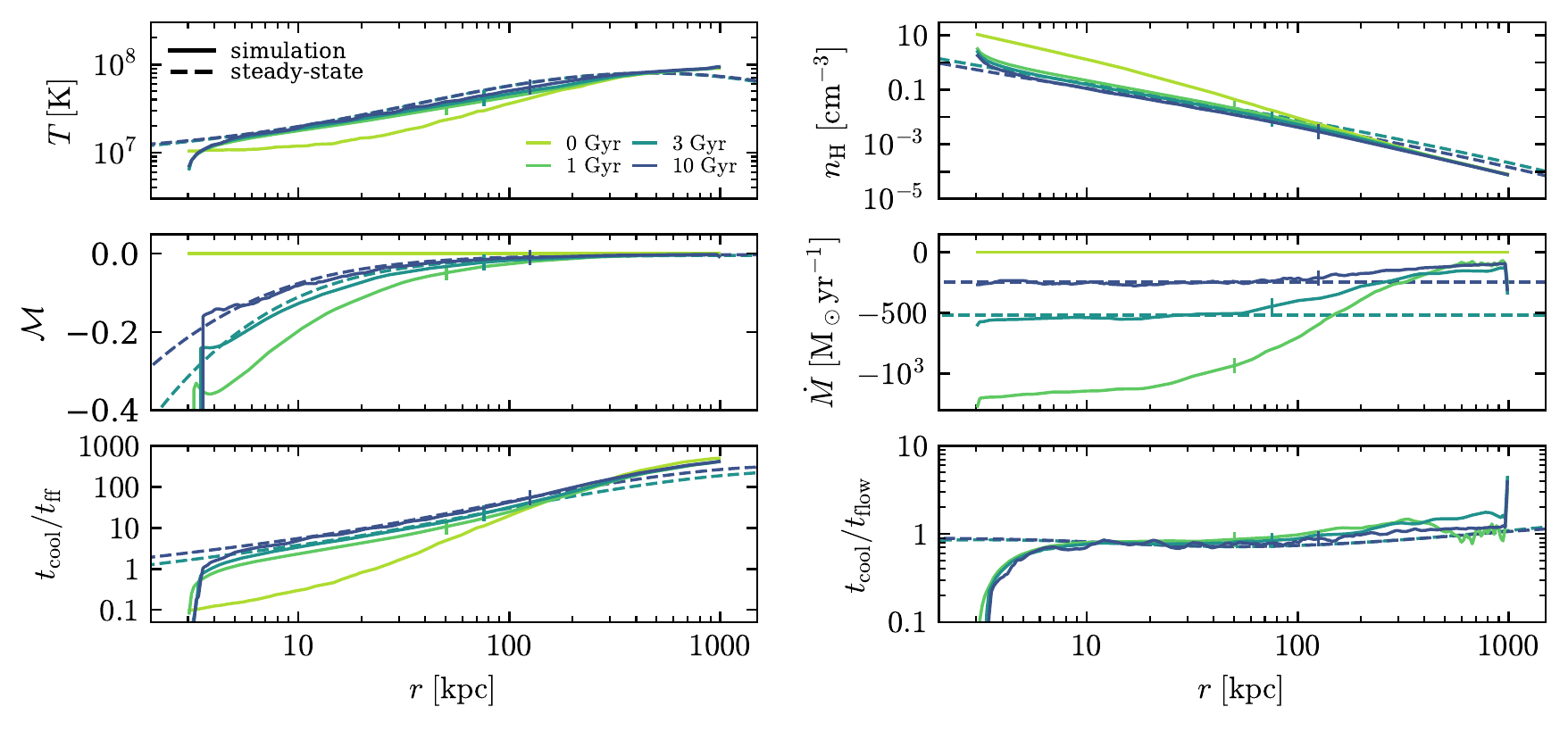}
\caption{
Spherically-averaged properties in the $10^{15}\msun$ simulation versus steady-state cooling flow solutions. 
The different panels show the temperature, density, Mach number, inflow rate, $\tcool/\tff$, and $\tcool/\tflow$. Colored solid lines plot spherically-averaged properties at different simulation times. Vertical ticks mark $\Rcool$, where the cooling time equals the simulation time.
Initial conditions are static, with $\mach=\Mdot=0$ (light green lines). 
Colored dashed lines plot two solutions to the steady-state equations as discussed in section~\ref{s:analytic}, corresponding to the $t=3\Gyr$ and $t=10\Gyr$ snapshots. The single free parameter $\Mdot$ of these solutions is set to equal $\Mdot$ near the inner boundary of the corresponding snapshot (see middle-right panel). The spherically-averaged properties of the simulation are well reproduced by the steady-state solutions, between $\approx6\kpc$ (a factor of $\approx2$ from the inner boundary) and out to $\approx\Rcool$.
}
\label{f:M15 sims}
\end{figure*}

We adopt a spherical-polar coordinate system for our simulations. Given that our initial and boundary conditions are spherically-symmetric, we do not expect deviations from spherical symmetry to span broad angular ranges, so we restrict our computational domain to $\pi/4 \leq \theta \leq 3 \pi / 4$ and $\pi/4 \leq \phi \leq 3 \pi / 4$, where $\theta$ is the polar angle and $\phi$ is the azimuthal angle. This enables us to reduce the computation expense of a given simulation and to avoid the small time steps that arise from the Courant-Friedrichs-Lewy (CFL) condition near the poles. In the radial direction our domain extends from $3-1000\kpc$ with logarithmic grid spacing. The outer radial boundary is chosen to be beyond $\Rcool$ (eqns.~\ref{e:Rcool Lstar} and \ref{e:Rcool cluster}), while the inner boundary is chosen to be within the galaxy scale for $\gtrsim10^{12}\msun$ halos (eqn.~\ref{e:Rhalf}). 
Our fiducial simulations have approximately 1:1 cell aspect ratios with $64$ cells in the angular directions and $240$ in the radial direction. This corresponds to cell widths ranging from $73\pc$ at the inner boundary, $720\pc$ at $R=30\kpc$, and $24\kpc$ at the outer boundary. Simulations with twice and half the resolution resulted in nearly indistinguishable radial profiles of the gas properties}, indicating that the simulations are well converged at our fiducial resolution.

\subsubsection{Boundary conditions}\label{s:boundary}

A standard radial outflow boundary condition, which imposes zero gradients, is insufficient for our problem because the cooling flow solution that develops has non-zero gradients. Thus, if the flow is subsonic at the inner boundary (i.e. if $\Rsonic < R_{\rm inner} = 3\kpc$) then a zero gradient boundary precludes the development of a steady-state solution. We therefore adopt boundary conditions that quadratically extrapolates the primitive variables into the ghost zones. In the polar and azimuthal directions we adopt periodic boundary conditions.

In practice, the inner radial boundary condition acts as a sink that allows gas to flow through it without introducing significant numerical artifacts. 
The outer boundary is beyond $\Rcool$ in all cases, so there are few changes near the outer boundary over the durations of the simulations, and the extrapolation boundary condition primarily serves to maintain equilibrium with the gravitational potential.

\subsubsection{Initial conditions}\label{s:ICs}

We run simulations with a $z=0$ UV background and a range of halo masses, $\Mhalo$, and use the appropriate $\vc(r)$ plotted in Fig.~\ref{f:vc and Lambda}. We start with an initially static halo ($v(r)=0$), with the density and temperature given by the hydrostatic relation
\begin{equation}\label{e:hydrostatic}
 \frac{\partial\ln P}{\partial \ln r} = -\gamma \frac{\vc^2}{\cs^2} ~.
\end{equation}
One needs also to specify the entropy profile and the total gas mass. As any bound hydrostatic solution has $\cs^2\sim\vc^2$, for simplicity we assume $\cs^2=\vc^2$, which yields an entropy profile $K\propto r^{\gamma(\gamma -1)} \sim r^{1.1}$ at radii where $\vc(r)$ is flat.  We checked that our conclusions are not sensitive to the exact choice of the ratio $\vc^2/\cs^2$, as long as it is of order unity. The total gas mass within $\Rvir$ is set to equal some fraction of baryon budget $\fb\Mhalo$, as detailed below. 

To verify that multiphase structure does not develop significantly in the simulation, as suggested by linear theory when $\tcool>\tff$ (e.g.~\citealt{BalbusSoker89}), we impose isobaric density perturbations in the initial conditions. The perturbations are generated so that the power is evenly distributed ($P_{\delta \rho} \propto k^0$) between modes satisfying $ 1 \leq k R_{\rm outer} / 2 \pi \leq 64 $ with an amplitude so that $\sigrho = 0.1$, where $R_{\rm outer}$ is the outer radius of the domain. Testing indicates that our results are not sensitive to the exact choices we made for the perturbations, except where noted below.

\begin{figure*}
 \includegraphics{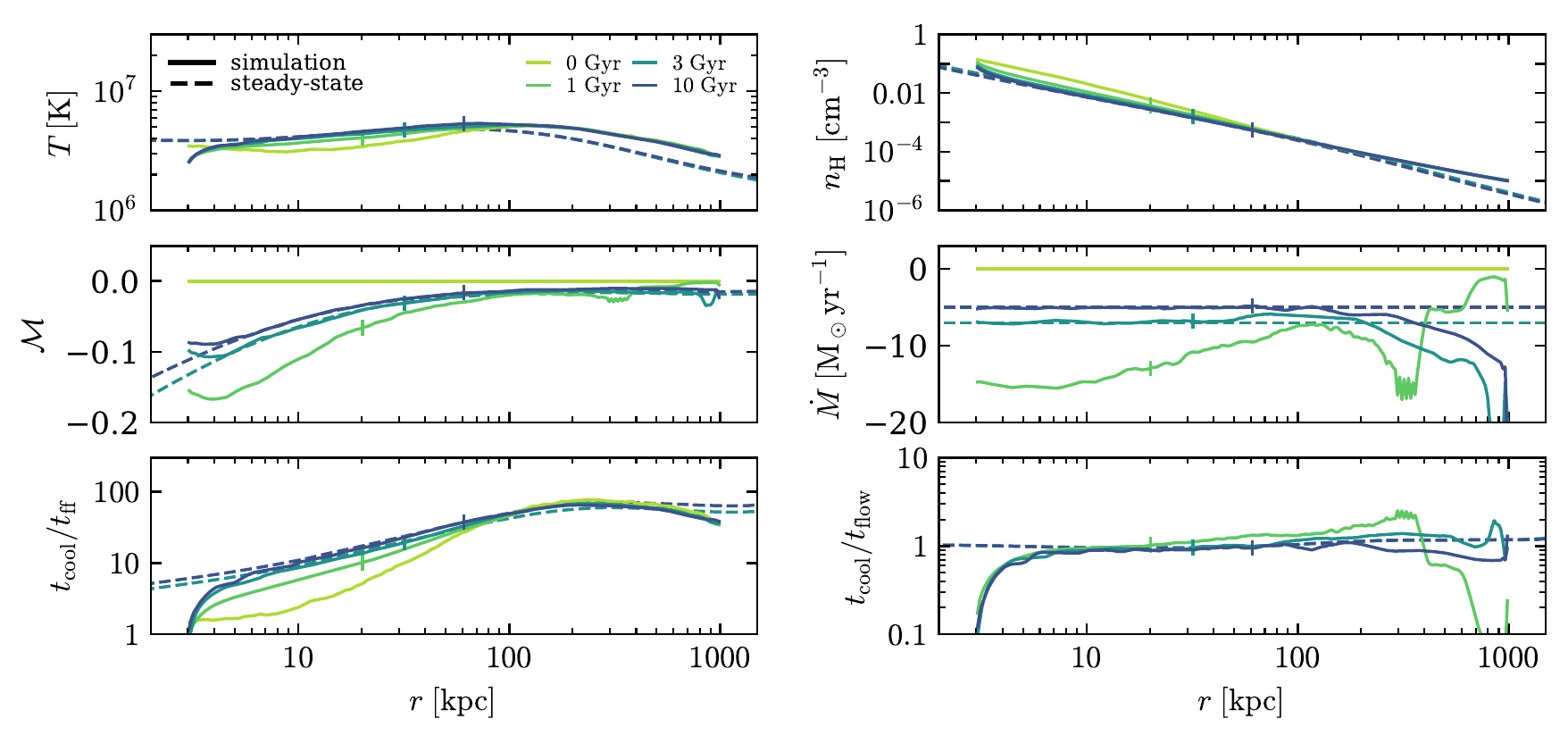}
\caption{
Similar to Fig.~\ref{f:M15 sims}, for the $10^{13}\msun$ simulation. 
The spherically-averaged properties of the simulation are well reproduced by the steady-state solutions, between $\approx6\kpc$ and $\approx\Rcool$.}
\label{f:M13 sims}
\end{figure*}

\subsection{Results}

Figure~\ref{f:snapshots} plots the results of four different simulations, with $\Mhalo=10^{15}\msun$ (left column), $10^{13}\msun$ (second column) and $10^{12}\msun$ (two right columns). The first three simulations initially have $M_{\rm gas} = \fb \Mhalo - \Mstar$ within $\Rvir$, while the simulation on the right starts with a gas mass equal to a third of this value. A metallicity of $\zsun/3$ is assumed throughout. In each column we show the temperature in units of $\Tvir$ (eqn.~\ref{e:Tvir}) and the radial velocity, at different simulation times $\tsim$ as noted in the panels. To project the snapshot onto the plotted $r-\theta$ plane we calculate the mass-weighted average of $T$ and $v$ over the simulated range in $\phi$. 
The initial hydrostatic conditions are plotted in the top row. Note how an inflow develops in all simulations.
Even though we seed perturbations in all simulations, strong deviations from spherical symmetry are apparent only at late times in the fiducial density $10^{12}\msun$ simulation, and near the inner boundary. As we show below, only at these regions and times the gas is supersonic and has $\tcool<\tff$, so multi-phase structure can develop. 

Solid colored lines in Figure~\ref{f:M15 sims} plot shell-averaged properties in the $10^{15}\msun$ simulation, as a function of radius and time. The panels show $T$ (top-left), $\nH$ (top-right), $\mach$ (middle-left), $\Mdot$ (middle-right), $\tcool/\tff$ (bottom-left), and $\tcool/\tflow$ (bottom-right). For each shell with radius $R$, we calculate mass-weighted averages for $T$, $\nH$, $\mach$, and $v$, while $\Mdot$ is the volume integral of $\rho v$ in the shell. The shell cooling time is calculated as
\begin{equation}\label{e:avg tcool}
 \tcool = \frac{\frac{3}{2}\int P\d\Omega }{\int \nH^2\Lambda\d\Omega }
\end{equation}
where $\d\Omega=\sin\theta\d\theta\d\phi$. Eqn.~(\ref{e:avg tcool}) implies that the shell cooling time is the ratio of the total thermal energy in the shell to the total luminosity of the shell. 

Vertical ticks in Fig.~\ref{f:M15 sims} mark the cooling radius $\Rcool$ at the different snapshots, defined via $\tcool(\Rcool,\tsim)=\tsim$. This radius mildly increases from $\Rcool=50\kpc$ at $\tsim=1\Gyr$ to $\Rcool=130\kpc$ at $\tsim=10\Gyr$. The top two panels demonstrate that within $\Rcool$, $T$ increases and $\nH$ decreases relative to their initial values (light green lines), while they are essentially unchanged at larger radii. 

The two dashed lines plot steady-state solutions with $\Rsonic=0.6\kpc$ (blue) and $0.3\kpc$ (purple), derived as discussed in section~\ref{s:analytic}. The values of $\Rsonic$ are chosen so $\Mdot$ in the steady-state solution equals $\Mdot$ in the $t=3\Gyr$ and $t=10\Gyr$ snapshots, respectively. Note that only the subsonic part of these two transonic solutions is within the simulated domain. The radial profiles of all plotted properties are well reproduced by the steady-state solutions, between $\approx6\kpc$ and $\approx\Rcool$. This result supports the main point of this paper, that cooling gaseous halos converge onto the steady-state cooling flow solutions described above, and that the entire solution is determined with a single free parameter.

\begin{figure*}
 \includegraphics{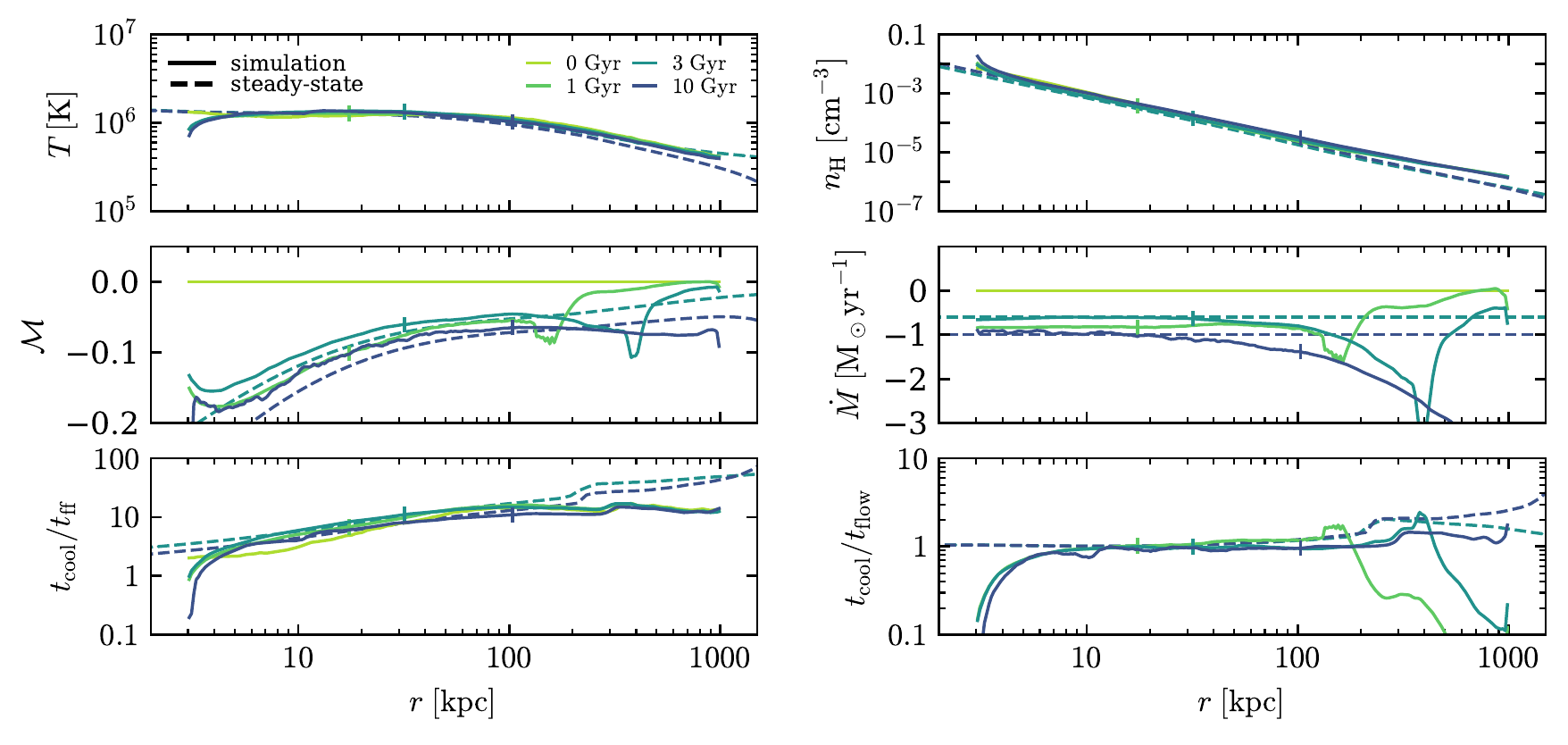}
\caption{
Similar to Fig.~\ref{f:M15 sims}, for the low-density $10^{12}\msun$ simulation. The spherically-averaged properties of the simulation are well reproduced by the steady-state solutions, between $\approx6\kpc$ and $\approx\Rcool$.
}
\label{f:M12 sims low density}
\end{figure*}

\begin{figure*}
 \includegraphics{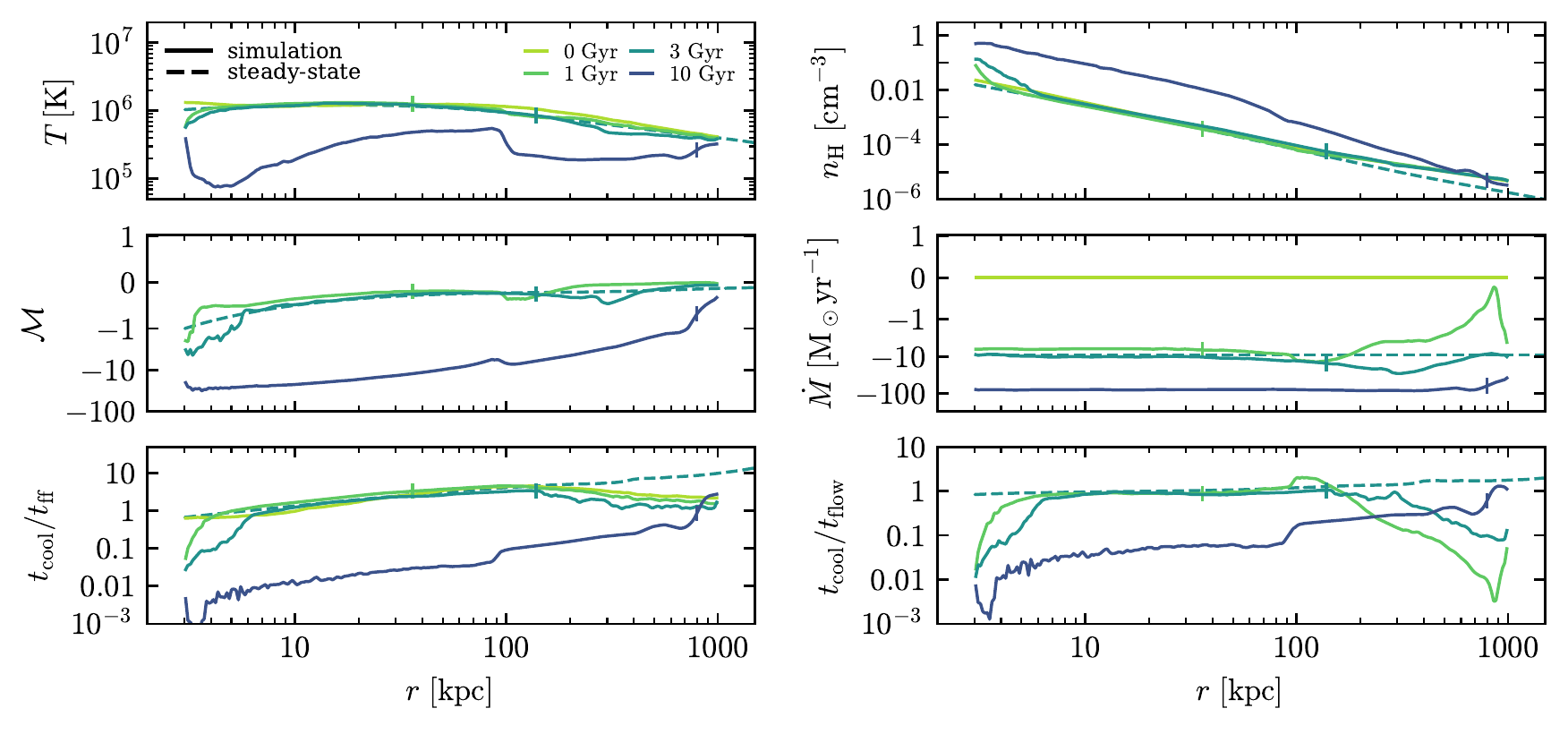}
\caption{
Similar to Fig.~\ref{f:M15 sims} for the fiducial density $10^{12}\msun$ simulation. 
Note that at $t=10\Gyr$ the cooling time is shorter than the free-fall time (bottom-left) and the flow is supersonic (middle-left). The cooling flow solutions are valid only for subsonic flows with $\tcool>\tff$. 
}
\label{f:M12 sims high density}
\end{figure*}

\begin{figure}
 \includegraphics{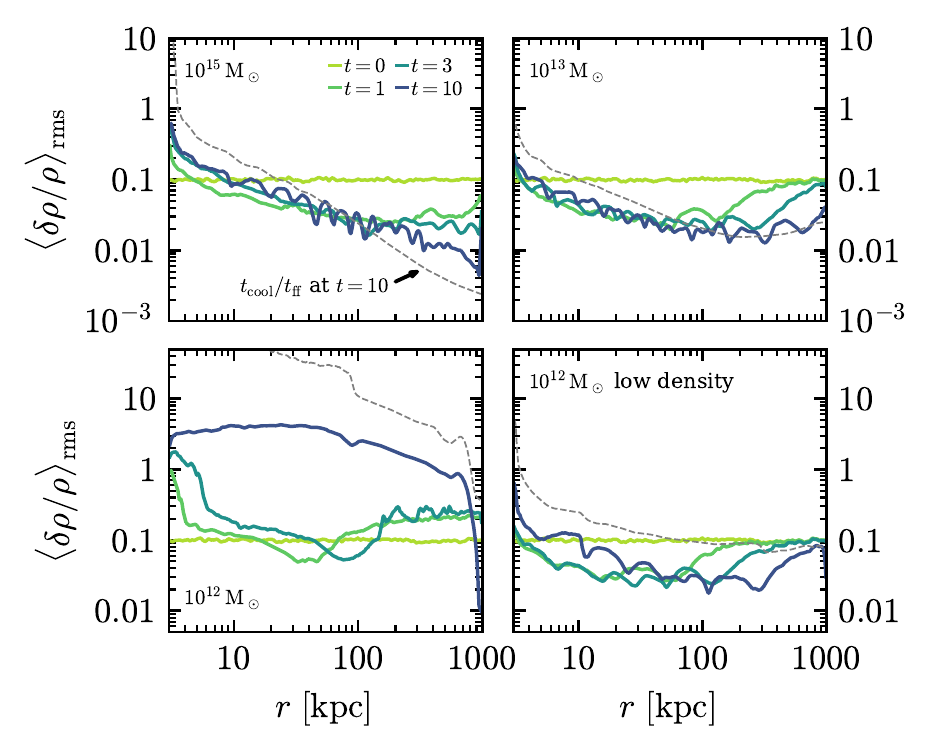}
\caption{Density fluctuations in the simulations. Each panel shows the normalized density dispersion in radial shells in a single simulation. Colored line denote snapshots at different times, as noted in Gyr in the top-left panel. 
Dashed gray lines plot $\tff/\tcool$ in the $t=10\Gyr$ snapshots.
We seed perturbations with $\sigrho=0.1$ in the initial conditions (light green lines). 
In most snapshots $\sigrho$ is smaller than unity throughout the domain, with the exception of the fiducial density $10^{12}\msun$ simulation at $t=10\Gyr$ (purple line in bottom-left panel). In this snapshot $\tcool<\tff$ and the flow is supersonic (Fig.~\ref{f:M12 sims high density}), compared to $\tcool>\tff$ and subsonic flows in the other snapshots (Figs.~\ref{f:M15 sims}--\ref{f:M12 sims high density}).
}
\label{f:sigma rho}
\end{figure}

The deviation of the radial profiles from the steady-state solutions within $\approx6\kpc$ is a boundary effect, since our boundary conditions are not exactly consistent with a steady-state solution (see section \ref{s:boundary}). Other choices of the inner boundary radius support this conclusion -- the radial profiles in the simulation deviate from a steady-state solution within a factor of $\approx2$ from the inner boundary. 

The lower-right panel in Fig.~\ref{f:M15 sims} shows that $\tcool\approx\tflow$ out to the outer boundary of the simulation, including radii larger than $\Rcool$ where $\tcool\gg\tsim$. 
This behavior is somewhat surprising given that a cooling flow is expected to form on a timescale $\tcool$. In Appendix~\ref{a:t_ratio} we demonstrate why the relation $\tcool\approx\tflow$ is established on a sound crossing timescale rather than a cooling timescale. The condition $\tcool\approx\tflow$ is though insufficient for a flow to lie on one of the steady-state solutions. The steady-state solutions also require a constant $\Mdot$ and the appropriate entropy profile (eqns.~\ref{e:mass1} and \ref{e:energy1}), which are both established on a timescale $\tcool$.  

What sets the free parameter $\Mdot$ in the simulation, which in Fig.~\ref{f:M15 sims} decreases from $520\msun\yr^{-1}$ at $t=3\Gyr$ to $240\msun\yr^{-1}$ at $t=10\Gyr$? Since the conditions beyond $\Rcool$ remain near their initial values, the free parameter must allow the cooling flow solution within $\Rcool$ to smoothly join the roughly static initial conditions beyond $\Rcool$. The value of $\Mdot(t)$ is hence set by the initial conditions at a radius $\Rcool(t)$. This relation was worked out analytically by \cite{Bertschinger89} for the self-similar case, who found that up to a factor of order unity
\begin{equation}\label{e:Mdot(t)}
 \Mdot(t) \approx 4\pi \Rcool^2\rho_0(\Rcool)\frac{\d \Rcool}{\d t}
\end{equation}
where $\rho_0(R)$ is the initial gas density at radius $R$. 
We find that $\Mdot$ near the inner boundary (at $R=20\kpc$) in the $10^{15}\msun$ simulation is consistent with eqn.~(\ref{e:Mdot(t)}) to within a factor of $1.5$. For comparison, the more naive estimate of $\Mdot(t)=M_{\rm gas}(<\Rcool(t))/t$ is larger by a factor of $\sim3$ then suggested by eqn.~(\ref{e:Mdot(t)}).
Note that if the flow is bounded on the outside by an accretion shock rather than by $\Rcool$ (see section~\ref{s:applicability}), the jump conditions at the shock will determine $\Mdot(t)$. 

Figures~\ref{f:M13 sims} and \ref{f:M12 sims low density} plot the spherically-averaged properties of the $10^{13}\msun$ simulation and the low-density $10^{12}\msun$ simulation, together with steady-state solutions corresponding to the $t=3\Gyr$ and $t=10\Gyr$ snapshots. As in Fig.~\ref{f:M15 sims}, the spherically-averaged properties of the simulation are well reproduced by the steady-state solutions, between $\approx6\kpc$ and out to $\approx\Rcool$. These simulations hence also support the main conclusion of this study, that within $\Rcool$ the halo gas settles on one of the single-parameter solutions derived in section~\ref{s:analytic}. The value of $\Mdot(t)$ near the inner boundary in these simulations is consistent with eqn.~(\ref{e:Mdot(t)}) to within a factor of two. 

Figure~\ref{f:M12 sims high density} plots average radial profiles in the $10^{12}\msun$ simulation with an initially closed baryon fraction (third column in Fig.~\ref{f:snapshots}). To increase the dynamic range we use a `symmetric log' axis in the $\mach$ and $\Mdot$ panels, where the y-axis is linear for absolute values smaller than unity and logarithmic for absolute values above unity. At $\tsim=3\Gyr$, the flow properties are consistent with the steady-state solution in the range $6\kpc < R < \Rcool$, as seen in the snapshots plotted in Figs.~\ref{f:M15 sims} -- \ref{f:M12 sims low density}. However, at $\tsim=10\Gyr$ the flow is supersonic, with $\tcool < \tff$ and a temperature significantly below virial. Both $\Rsonic$ and $\Rcool$ in this snapshot are at $800\kpc$. The steady-state solutions derived in the previous section are thus invalid if $\tcool<\tff$. 
The transition between a cooling flow and a supersonic inflow as seen in this simulation is discussed in Paper II.

Figure~\ref{f:sigma rho} plots the r.m.s. density dispersion $\sigrho$ in radial shells in the four simulations. Each panel corresponds to a different simulation, while each colored line corresponds to a different snapshot as noted in the legend in the top-right panel. The profiles of $\tff/\tcool$ in the $t=10\Gyr$ snapshots are plotted with dashed grey lines. In the $10^{15}\msun$ simulation (top-left panel), by $t=1\Gyr$ the amplitude of the perturbations has decreased from their inital value of $\sigrho=0.1$ (light green lines). Specifically, at $t=1$, $3$ and $10\Gyr$ the relative fluctuations are significantly below unity, indicating that multi-phase structure has not developed. 
A similar behavior is apparent in the $10^{13}\msun$ and low density $10^{12}\msun$ simulations (right panels) except at small $t$ and large $R$ where the initial perturbations remain. 
Significant fluctuations where $\sigrho$ is larger than unity are apparent only in the fiducial density $10^{12}\msun$ simulation at $t=10\Gyr$, in which $\tff/\tcool>1$  and the flow is supersonic. These fluctuations are also apparent in the corresponding temperature panel in Fig.~\ref{f:snapshots}. Figure~\ref{f:sigma rho} thus demonstrates that cooling flows do not develop into a multiphase CGM, as long as $\tcool > \tff$.

We emphasize that the latter conclusion applies only to perturbations with an initial amplitude significantly lower than unity. Stronger perturbations seeded e.g.~by the wakes of satellite galaxies are likely to persist in the halo, and may affect the evolution of the ambient medium even if it has $\tcool\gtrsim\tff$ (e.g.~\citealt{Sharma+12a,Choudhury+19}). 

The decrease in amplitude of the perturbations from their initial value occurs roughly on a free-fall time, which increases with decreasing halo mass at fixed $r$. After this initial phase, in the three simulations which do not collapse the amplitude of the perturbations is within a factor of $2-3$ of $\tff/\tcool$ (compare the purple lines with the dashed gray lines). We find an almost identical final perturbation amplitude in simulations initialized with a higher $\sigrho=0.3$, though simulations with a lower initial $\sigrho=0.01$ have a final $\sigrho\ll\tff/\tcool$. A similar saturation of density fluctuations at an amplitude $\approx\tff/\tcool$ has previously been seen in simulations of thermally balanced atmospheres, and attributed to dissipation of perturbations of this amplitude via non-linear mode coupling (\citealt{McCourt+12, Voit+17}).

\section{Comparison with Observations}\label{s:observations}

\begin{figure*}
 \includegraphics{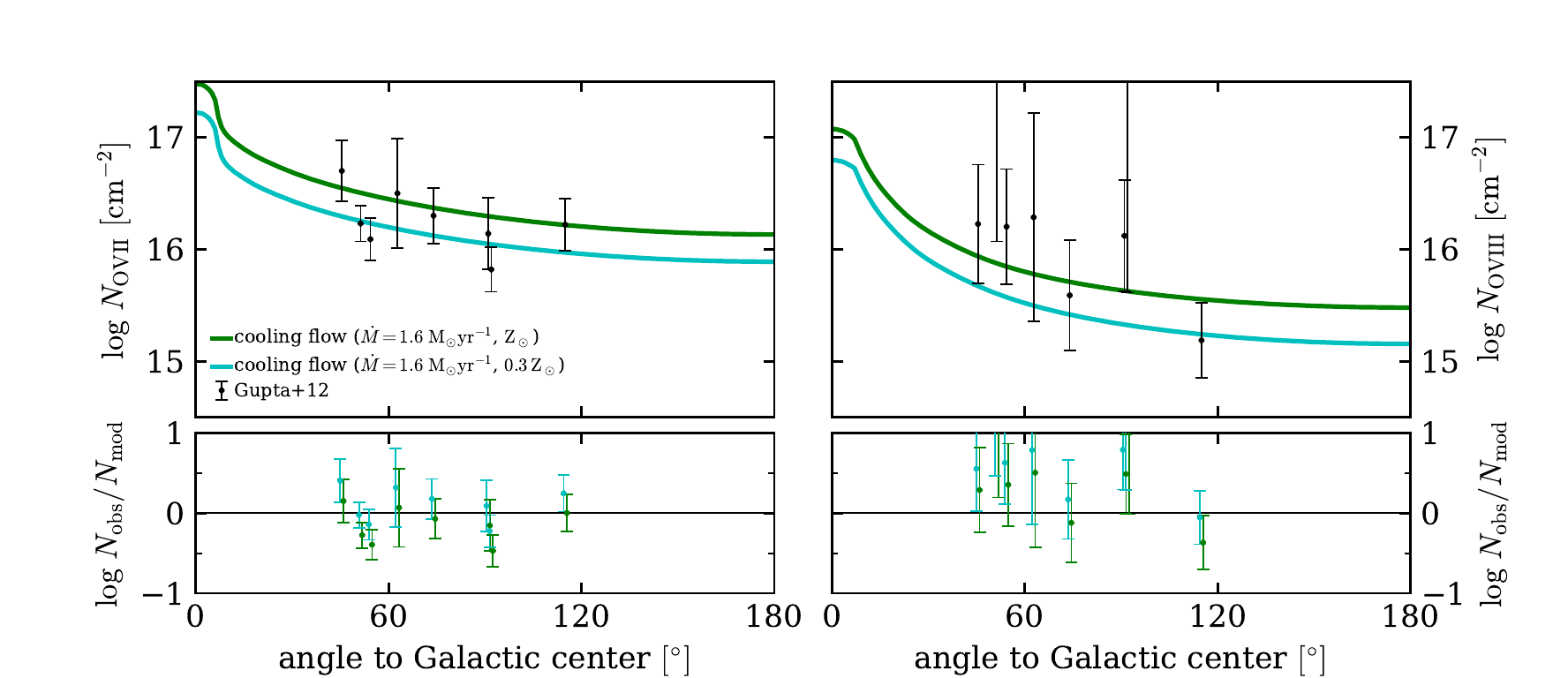}
\caption{Comparison of cooling flow predictions with \ion{O}{VII} (left) and \ion{O}{VIII} (right) absorption in the Milky Way halo. 
\textbf{(Top panels)} Colored lines plot the cooling flow predictions as a function of line-of-sight angle to the Galactic center, for two assumed halo gas metallicities. 
The free parameter of the solutions is set so $\Mdot$ equals the MW star formation rate of $1.6\msun\yr^{-1}$. Error bars denote measured columns from Gupta et al.~(2012). 
\textbf{(Bottom panels)}
The difference between the data and the cooling flow predictions, colored by the metallicity of the cooling flow solutions. The error bars are slightly offset horizontally for clarity. The cooling flow solutions are consistent with the observed MW X-ray absorption, with no free parameters beyond the uncertainty in metallicity.
}
\label{f:MW absorption}
\end{figure*}

In this section we compare observables of cooling flow solutions with observational constraints of halo gas at low redshift. 

\subsection{OVII and OVIII absorption in the Milky Way}\label{s:MW absorption}

The expected column of an ion $X^{i+}$ in the halo can be derived via 
\begin{equation}\label{e:Nion}
 N_{X^{i+}} = \frac{\rm X}{\rm H}\int \nH f_{\rm X^{i+}}(T,\nH)\d s ~,
\end{equation}
where $N_{X^{i+}}$ is the ion column, ${\rm X}/{\rm H}$ is the abundance of element ${\rm X}$ relative to hydrogen,  $f_{\rm X^{i+}}$ is the fraction of X particles in the $i$-th ionization state, and $\d s$ is the line-of-sight element. For $f_{\rm X^{i+}}$ we use the ionization equilibrium calculations in the \trident\ package (\citealt{Hummels+17}), which used \cloudy\ (\citealt{Ferland+13}) and assumed the gas is exposed to a HM12 background. As in section \ref{s:basic eqs}, we assume the actual background has twice the intensity deduced by HM12.

We calculate two cooling flow solutions with a metallicity of either $0.3\zsun$ or $\zsun$, $\Mdot=1.65\msun\yr^{-1}$ equal to the SFR of the Galaxy, $\Mhalo=1.3\cdot10^{12}\msun$, and $\Mstar=6.3\cdot10^{10}\msun$. All estimates for Galactic parameters are taken from \cite{BlandHawthornGerhard16}. 
The choice of $\Mdot={\rm SFR}$ is motivated by the assumption that all fuel for star formation is provided by the hot CGM, though lower $\Mdot$ are also possible in the context of a cooling flow CGM, if some of the fuel is supplied by stellar mass loss (\citealt{LeitnerKravtsov11}).
The values of the metallicity are chosen to bracket the possible range, where the lower limit was deduced by \cite{MillerBregman13, MillerBregman15}, who divided the observed $\Novii$ by the upper limit on the dispersion measure towards the LMC (see section \ref{s:DM} below), while the upper limit is the ISM metallicity.

Using $T(r)$ and $\nH(r)$ from the cooling flow solutions (where $r$ is the galactocentric radius), we integrate eqn.~(\ref{e:Nion}) along different sightlines through the MW halo. For a given solution, the predicted columns depend on the angle of the sightline relative to the Galactic center.  We end the integration at $\Rcool=130\kpc$ (for the $0.3\zsun$ solution) or at $\Rcool=180\kpc$ ($\zsun$), though since more than 85\%\ of the absorption originates within the inner $\sim40\kpc$, the outer end of the integration does not significantly affect the predicted columns. The predicted $\Novii$ and $\Noviii$ are shown in the top-left and top-right panels of Figure~\ref{f:MW absorption}, respectively.

The observed $N_{\rm OVII}$ plotted in the top-left panel of Fig.~\ref{f:MW absorption} are taken from \citeauthor{Gupta+12} (2012, hereafter G12), who presented a sample of eight high S/N sightlines with detections in both the K$\alpha$ (21.6\AA) and K$\beta$ (18.63\AA) absorption lines. 
For comparison, \cite{Faerman+17} deduced a similar median $\Novii=1.4\cdot10^{16}\cm^{-2}$ based on the sample of \cite{Fang+15}, which included 43 sightlines with lower S/N than G12. The $N_{\rm OVII}$ measurements in \cite{Fang+15} are based solely on the K$\alpha$ line and typically have $\gtrsim$$1\,{\rm dex}$ uncertainties, so we do not show them individually. The bottom-left panel shows the difference between the observations and the predictions of the cooling flow solutions.
The cooling flow solution with $Z=\zsun/3$ is consistent with the G12 observations (reduced $\chi^2=0.8$) while the $Z=\zsun$ solution is marginally consistent (reduced $\chi^2=1.7$). All solutions with metallicities $0.3-0.7\zsun$ yield an acceptable fit with a reduced $\chi^2$ lower than unity.

The observed $N_{\rm OVIII}$ shown in the top-right panel of Fig.~\ref{f:MW absorption} are also based on the G12 data. We calculate $N_{\rm OVIII}$ using a curve-of-growth analysis, based on the \ion{O}{VIII} equivalent widths and the velocity width parameter $b$ measured for \ion{O}{VII}, which is justified if the two ions originate in the same gas (see similar calculation in \citealt{Faerman+17}). 
Comparison of these measurements with the models yields a reduced $\chi^2=0.7$ for the $\zsun$ model and a reduced $\chi^2=1.1$ for the $\zsun/3$ model.  Combining the $N_{\rm OVII}$ and $N_{\rm OVIII}$ observations, the cooling flow solutions yield acceptable fits for metallicities in the range $0.3-0.8\zsun$. 
Fig.~\ref{f:MW absorption} thus demonstrates that the cooling flow solutions are consistent with the observed  $N_{\rm OVII}$ and $N_{\rm OVIII}$. We emphasize that there are no free parameters in these solutions beyond the uncertainty in $Z$.

The cooling flow solutions used to calculate the predicted columns in Fig.~\ref{f:MW absorption} assume a mass inflow rate equal to the SFR in the Milky Way. We also calculate solutions with different values of $\Mdot$ and compare them to the observational data. We find an acceptable fit for $1.6<\Mdot<4\msun\yr^{-1}$ if $Z=\zsun/3$, and for $0.2<\Mdot<1.3\msun\yr^{-1}$ if $Z=\zsun$. Therefore, if the MW halo gas forms a cooling flow, then the observed X-ray ion columns allow for mass inflow rates in the range $0.2 - 4\msun\yr^{-1}$, or equivalently $0.12 - 2.4$ times the SFR.

\subsection{OVII and OVIII emission in the Milky Way}\label{s:MW emission}

\begin{figure}
 \includegraphics{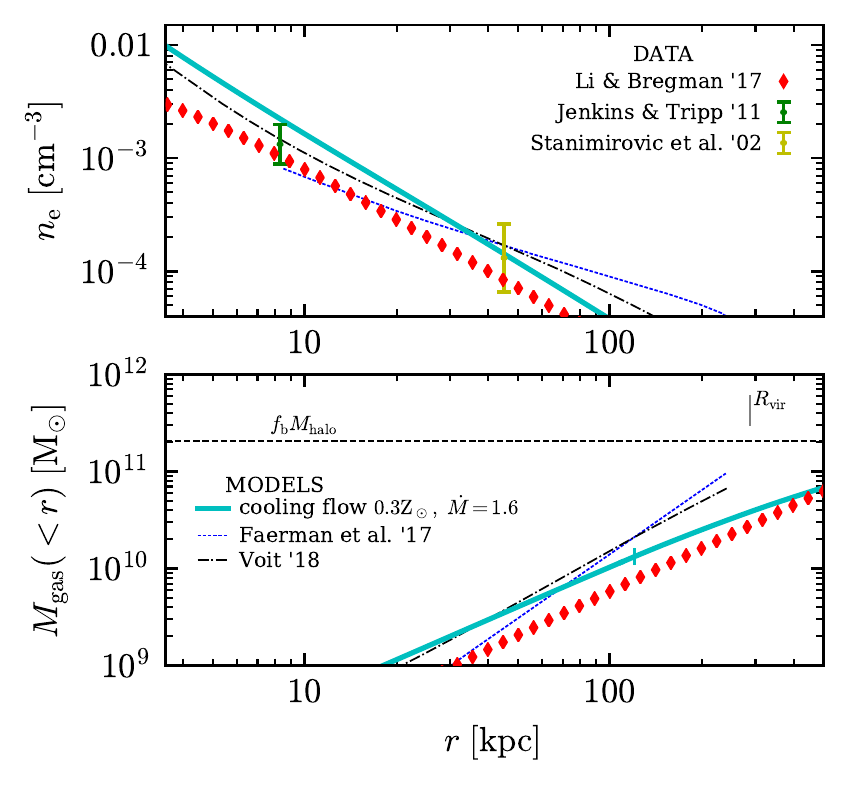}
\caption{Hot gas density and mass profiles in cooling flow solutions versus observational constraints. 
The cyan line plots a cooling flow solution with $\Mdot$ equal to the Milky Way SFR and $Z=0.3\zsun$. 
Diamonds denote the fit profiles of LB17, constrained via the observed \ion{O}{VII} and \ion{O}{VIII} emission along 648 sightlines through the MW halo, also assuming $Z=0.3\zsun$. 
The radial density slope in cooling flows is consistent with the observationally-constrained LB17 profile at $\gtrsim10\kpc$, while the density normalization is a factor of $\approx2$ higher. 
Error bars denote hot gas density estimates based on  thermal pressure measurements in the local ISM and Magellanic Stream. 
In the bottom panel a small tick on the cooling flow solution marks $\Rcool$, beyond which the solution is extrapolated. The implied total hot gas mass within $\Rvir$ is $17\%$ of the halo baryon budget. 
The blue dashed-dotted lines plot the Faerman et al.~(2017) model which assumes feedback heating of the halo gas. This model has a flatter density profile than cooling flows and a closed baryon fraction within $\Rvir$. 
The black dashed-dotted lines plot the Voit (2018) model in which the gas has $\tcool=10\tff$.  }
\label{f:MW density}
\end{figure}

\begin{figure*}
 \includegraphics{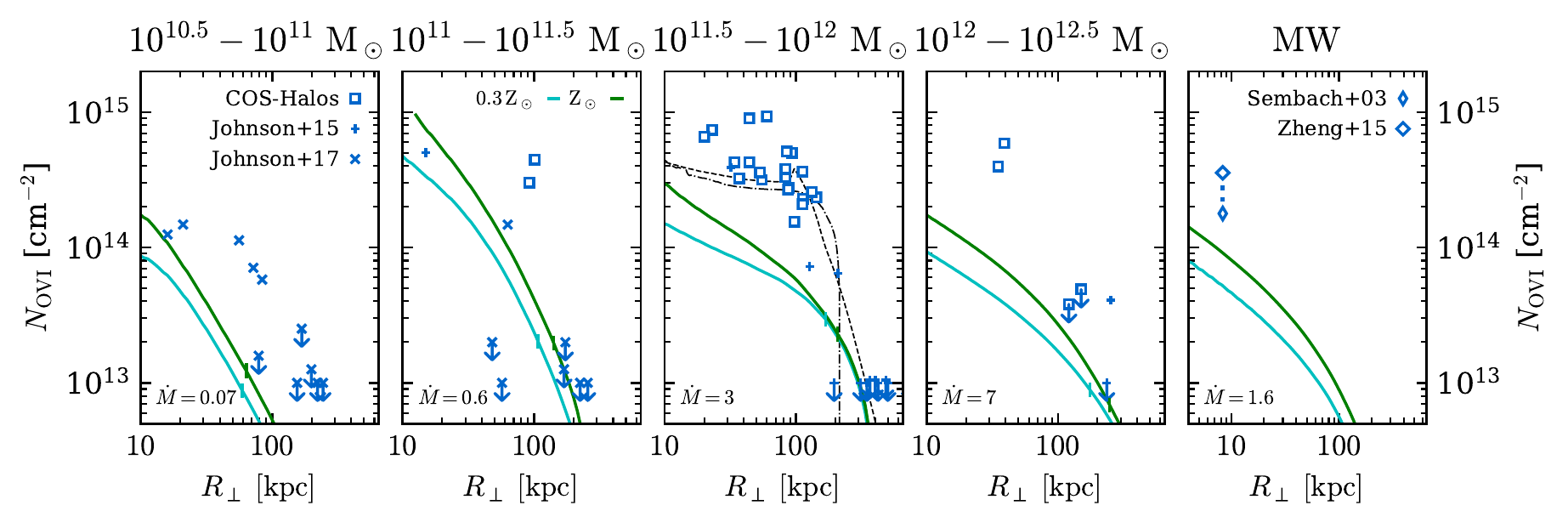}
\caption{
Comparison of cooling flow solutions with \ion{O}{VI} absorption columns around star-forming galaxies at $z\sim0.2$ (four left panels) and around the Milky Way (right panel). External galaxies are grouped by halo mass (noted on top), estimated from the stellar mass of the central galaxy. Cooling flow predictions are plotted for $Z=0.3\zsun$ (cyan) and $Z=\zsun$ (green), assuming $\Mdot$ equals the median SFR in each $\Mhalo$ bin in the observed sample ($\Mdot$ noted in $\msun\yr^{-1}$ in the bottom-left of each panel). 
In the MW panel the lower marker plots the average $\Novi$ from Sembach et al.~(2003), multiplied by two to mimic an external sightline, while the upper marker includes the correction from Zheng et al.~(2015) due to confusion with absorption in the Galactic disk. The cooling flow solutions for the MW halo are the same solutions as used in Fig.~\ref{f:MW absorption}. Cooling flows typically underpredict the observed $N_{\rm OVI}$ by a factor of $\sim5$. In the middle panel we plot $\Novi$ of two suggested models which roughly fit the \ion{O}{VI} observations, while preserving the good fit of the cooling flow solutions to the X-ray observations (Figs.~\ref{f:MW absorption}--\ref{f:MW density}). The black dash-dotted line denotes a model with a heating mechanism which preferentially affects the outer halo,
while the dashed line denotes a model in which cool photoionized gas shocks at $100\kpc$, and subsequently forms a cooling flow at smaller radii (see section~\ref{s:discrepancy} for details). 
}
\label{f:OVI}
\end{figure*}

\citeauthor{LiBregman17} (2017, hereafter LB17) derived the density profile of the Milky Way halo gas based on X-ray observations of \ion{O}{VII} and \ion{O}{VIII} emission lines, using 648 sightlines from the sample of \cite{HenleyShelton12,HenleyShelton13}. In order to derive the density profile they assumed a constant temperature of $2\cdot10^6\K$, similar to the cooling flow solutions for the MW halo discussed in the previous section, in which the temperature can be approximated as $T/10^6\K\approx 1.6-0.5\log(R/10\kpc)$. The observationally-constrained density profile of LB17 can thus be directly compared to the density profile predicted by cooling flow solutions. 

The most detailed model in LB17 (model `9'), 
which accounted for radiation transfer effects, included a disk component and allowed for rotation of the hot gas, yielded an electron density profile at large radii of 
\begin{equation}\label{e:LB17}
 n_{\rm e}(r\gg2.5\kpc) = (0.028\pm0.003) \left(\frac{R}{\rm kpc}\right)^{-1.53 \pm 0.02} ~.
\end{equation}
This density profile  was constrained using the \ion{O}{VII} emission line triplet near 22\AA. LB17 also constrained their model based on the \ion{O}{VII} and \ion{O}{VIII} emission lines near 19\AA\ (their model `3'), which yielded a profile consistent with eqn.~(\ref{e:LB17}), though with larger errors. 
The top panel of Figure~\ref{f:MW density} plots the LB17 density profile versus the MW cooling flow solution with $\Mdot={\rm SFR}$ and $Z=0.3\zsun$. LB17 also assumed $Z=0.3\zsun$, though since the density in cooling flows and in LB17 scale similarly with the assumed $Z$, the assumed $Z$ does not affect the comparison. 
Fig.~\ref{f:MW density} shows that the slope of the LB17 model and the cooling flow solutions are almost identical at radii $\gtrsim10\kpc$, as also evident from comparing eqn.~(\ref{e:LB17}) with eqn.~(\ref{e:nH Lstar}). As the density slope is a robust prediction of the cooling flows, the LB17 result supports the hypothesis that the MW halo gas beyond $\gtrsim10\kpc$ forms a cooling flow. 
At smaller radii the LB17 profile flattens while the cooling flow profile remains steep, though we do not expect the cooling flow to apply at such small radii due to angular momentum and deviations of the potential from spherical symmetry. 

Fig.~\ref{f:MW density} shows that the normalization of the LB17 and cooling flow profiles differ by a factor of $\approx 2$. Roughly $30\%$ of this difference can be accounted for by the higher temperature assumed in LB17 compared to the temperatures in the cooling flow solution, which yields a higher emissivity and hence a lower density for a given line emission. The remaining difference may suggest that the actual mass inflow rate is a factor of $\approx 2$ lower than the value of ${\rm SFR}=1.65\msun\yr^{-1}$ assumed in the cooling flow solution. 

To produce a good fit to the observations, LB17 assumed an uncertainty of 2.1 L.U. in their model for the \ion{O}{VII} emission ($\approx40\%$ of the median \ion{O}{VII} intensity). They attributed this uncertainty to unaccounted variations in the emission from either the Local Bubble or the halo gas. If these emission variations are not due to the Local Bubble but arise in halo gas, they could indicate density variations of order $\lesssim 20\%$ in the cooling flow relative to spherical symmetry (since emission scales as $\nH^2$). 

In the bottom panel of Fig.~\ref{f:MW density} we show the total hot gas mass in the MW halo implied by the cooling flow solutions. The solutions are extended beyond $\Rcool$ (marked by a tick) assuming a hydrostatic pressure profile with $K\propto R$, though we note that this extension is uncertain. The implied total gas mass within $\Rvir$ is $17\%$ of the halo baryon budget. Extending the profile beyond $\Rcool$ with a shallower entropy profile yields a somewhat higher gas mass (e.g.~$K\propto R^{0.5}$ yields a gas mass of $22\%$ of the baryon budget), while assuming a metallicity higher than $0.3\zsun$ decreases the implied gas mass by $\approx (Z/0.3\zsun)^{-1/2}$. 
A similar low baryon mass was derived by \cite{Bregman+18}, who deduced the total hot gas mass by extrapolating the LB17 gas density profile out to $\Rvir$ (marked with diamonds in Fig.~\ref{f:MW density}).

\subsection{Dispersion measure towards the LMC}\label{s:DM}

Using measurements of the dispersion measure towards pulsars in the Large Magellanic Cloud (LMC), \cite{AndersonBregman10} estimated an  electron column of $23\cm^{-3}\pc$ in the halo, after subtracting an estimated contribution of $47\cm^{-3}\pc$ from the Galactic disk and assuming a negligible contribution from gas in the LMC. For comparison, the cooling flow solution with $\Mdot={\rm SFR}$ and $Z=0.3\zsun$ predicts a dispersion measure between $8.3$ and $50\kpc$ of $37\cm^{-3}\pc$, while the $Z=\zsun$ solution predicts $22\cm^{-3}\pc$. The predicted and observed values are comparable, supporting the cooling flow solution for hot gas in the MW halo. The somewhat lower values suggested by the dispersion measure observations, especially if $Z$ is relatively low or if the LMC contribution is non-negligible, may suggest $\Mdot$ is actually somewhat smaller than the SFR, as also suggested by the comparison of the cooling flow solution with the LB17 profile in Fig.~\ref{f:MW density}.

\subsection{OVI absorption}\label{s:OVI}

Figure~\ref{f:OVI} compares the prediction of cooling flow solutions with observations of \ion{O}{VI} absorption around $z\sim0.2$ star-forming galaxies and around the Milky Way. Blue markers in the four left panels denote $\Novi$ measurements from \cite{Werk+13}, \cite{Johnson+15}, and \cite{Johnson+17}. The galaxies are grouped by $\Mhalo$ as noted at the top of each panel, where $\Mhalo$ is estimated from the stellar masses of the central galaxies noted in the papers, using the \cite{Behroozi+18} relation\footnote{One object, J1435+3604\_68\_12, has $\Mhalo=10^{13}\msun$ and is not shown. \ion{O}{VI} is not detected in this object.}. For each $\Mhalo$ bin, we calculate cooling flow models with the median $z=0.2$, either $Z=0.3\zsun$ or $Z=\zsun$, and $\Mdot$ equal to the median SFR in the observed galaxies, taken from \cite{Werk+13}. In the left panel where all galaxies do not have a published SFR estimate we assume $\Mdot=0.07\msun\yr^{-1}$, the average SFR for galaxies with the median $\Mstar=1.6\cdot10^8\msun$ in this bin (\citealt{Behroozi+18}). 
In the right panel the lower marker shows the average of observations along MW sightlines from \cite{Sembach+03}, multiplied by two to mimic an external galaxy sightline as in the other panels. Due to confusion with absorption in the Galactic disk, the \cite{Sembach+03}\ observations do not include \ion{O}{VI}\ absorption at local standard of rest velocities $\left|v_{\rm LSR}\right|<100\kms$, which suggests $\Novi$ are actually a factor of $\sim2$ higher (\citealt{Zheng+15}). The corrected average $\Novi$ is noted with the upper marker. As mentioned by \cite{Zheng+15}, the corrected $\Novi$ in the Milky Way are similar to columns observed in other star-forming galaxies with the same mass, suggesting a similar physical origin. 
The cooling flow solutions in the right panel are the same solutions used in Fig.~\ref{f:MW absorption}.

The predicted ion columns are calculated from the cooling flow solutions using eqn.~(\ref{e:Nion}), integrated along sightlines with different impact parameters $R_{\perp}$. We integrate out to $2\Rvir$ rather than to $\Rcool$ to avoid an unphysical break in the predicted $\Novi(R_\perp)$ profile near $\Rcool$, though in all cases the contribution to $\Novi$ of gas beyond $\Rcool$ is small for sightlines with $R_\perp<\Rcool$. 
The cooling flows solutions underpredict the observed $\Novi$ around most galaxies, typically by a factor of $\sim 5$. A similar conclusion arises from the MW observations (right panel) where the solutions underpredict the average \cite{Sembach+03} observations by a factor of $1.8-3.2$, and the corrected observations from \cite{Zheng+15} by a factor of $3.5-6.5$. This failure in reproducing the observed $\Novi$ is in contrast with the cooling flow solutions success in reproducing the observed \ion{O}{VII} and \ion{O}{VIII} absorption (Fig.~\ref{f:MW absorption}), the density slope suggested by the observed \ion{O}{VII} and \ion{O}{VIII} emission (Fig.~\ref{f:MW density}), and the dispersion measure towards the LMC (section~\ref{s:DM}).
In the discussion we suggest possible resolutions to this apparent discrepancy, in which either the \ion{O}{VI}-bearing outer halo is preferentially heated, or alternatively the cooling flow extends only out to an accretion shock at $\lesssim100\kpc$ and \ion{O}{VI} predominantly traces cool gas beyond the shock.

The predicted $\Novi$ in cooling flows scales roughly as $\Novi \propto \nH f_{\text{\sc O\,vi}} \propto \Mdot^{1/2}$, since $\nH\propto\Mdot^{1/2}$ (eqn.~\ref{e:nH Lstar}) while $f_{\text{\sc O\,vi}}$ which depends mainly on $T$ is roughly independent of $\Mdot$. Hence, in principle a cooling flow with $\Mdot\sim25\,{\rm SFR}$ could reproduce the \ion{O}{VI} observations, though such a model would have $\tcool<\tff$ and hence would be unstable. It has been shown that any model which assumes \ion{O}{VI} traces radiatively cooling gas requires $\Mdot\gg{\rm SFR}$ (\citealt{MathewsProchaska17,Faerman+17,McQuinnWerk18, Stern+18}).

\subsection{Cool clouds}\label{s:cool clouds}

Cool gas clouds ($\sim10^4\K$) are routinely observed in dark matter halos via their low-ion (e.g.~\ion{Mg}{II}, \ion{C}{II}, \ion{Si}{II}) and \ion{H}{I} absorption (see review by \citealt{Tumlinson+17}). Around low-redshift $\sim$$L^*$ galaxies, such absorption features appear in $\gtrsim50\%$ of sightlines to background quasars with impact parameter out to $\approx0.5\Rvir$ (\citealt{Werk+13,LiangChen14}, see also figure~11 in \citealt{Stern+18}). 
The prevalence of cool gas in halos is a challenge for the validity of subsonic cooling flow solutions, in which the halo gas is expected to be predominantly single-phased. This expectation is evident in Fig.~\ref{f:sigma rho}, which demonstrates that in simulations seeded with small amplitude perturbations ($\sigrho=0.1$, see section~\ref{s:ICs}) the density dispersion remains smaller than unity in all snapshots where $\tcool>\tff$, consistent with results based on linear perturbation theory (\citealt{Malagoli+87, BalbusSoker89}).

It is however important to note that due to projection effects, the large observed covering factor of cool clouds does not necessarily imply that they are widespread throughout the halo. Cool clouds clustered in a small fraction of the halo volume near `local' disturbances which produce non-linear perturbations, such as in the extended disk, along collimated outflows, or near satellites and the material stripped from them (e.g.\ the Magellanic Stream), could in principle have a large area covering factor. 
This association of cool clouds with specific locations in the halo is suggested by the tendency of \ion{Mg}{II} clouds to align with either the minor axis or the major axis of the galaxy (\citealt{Bouche+12,Kacprzak+12,Nielsen+15,Martin+19}).
It is thus a prediction of the cooling flow scenario that the observed cool clouds are limited to a small fraction of the halo volume. 
Such a scenario can be seen in the FIRE simulations where a large fraction of the cool gas in $\gtrsim10^{12}\msun$ halos is associated with satellite galaxies and their outflows (\citealt{FaucherGiguere15,FaucherGiguere16}), and in the simulations of \cite{Hummels+18}, where cool clouds form only near the galaxy or near an inflowing filament.

In regions where cool clouds do form, they can be used as a rough barometer of the ambient virial-temperature gas by assuming pressure equilibrium between the cool and hot phases, as initially done by \cite{Spitzer56}. In the top panel of Fig.~\ref{f:MW density} we plot the hot gas density implied by the pressure of clouds in the Magellanic stream derived by \cite{Stanimirovic+02}, who used 21cm observations and deduced $P/k=300\cm^{-3}\K$ assuming a distance of $45\kpc$. We convert their pressure estimate to a density estimate using $T(45\kpc)=1.3\cdot 10^{6}\K$ in the cooling flow solution. We estimate an error of $0.3\,{\rm dex}$ on the \cite{Stanimirovic+02} estimate based on the dispersion between different Magellanic stream clouds (see their figure~16). This pressure estimate is consistent with the cooling flow solution.  

Also plotted in Fig.~\ref{f:MW density} is the hot gas density implied by the pressure in the local interstellar medium, as measured by
\cite{JenkinsTripp11}. They used observations of \ion{C}{I} absorption and deduced a pressure of $P/k=3800\cm^{-3}\K$ with a dispersion of $0.175\,{\rm dex}$, which we convert to a density estimate using $T(8.3\kpc)=1.6\cdot 10^{6}\K$ in the cooling flow solution. This pressure estimate is marginally consistent with the $0.3\zsun$ cooling flow solution. The median value of \cite{JenkinsTripp11} is equal to the prediction of the $Z=\zsun$ cooling flow solution (not shown). 

The pressure of cool clouds in halos can also be constrained via photoionization modelling of low-ionization UV absorption features. In the COS-Halos sample of low-redshift star-forming galaxies, \cite{Werk+14} and \cite{McQuinnWerk18} deduced a gas pressure 
which is a factor of $\gtrsim10$ lower than expected in a gaseous
halo with a closed baryon fraction. Broadly consistent results were obtained on a similar sample by \citeauthor{Keeney+17} (2017, see also \citealt{Voit+19}). Though these estimates are subject to several systematic uncertainties (\citealt{Stern+16, Chen+17}), taking these pressure estimates at face value suggests a hot gas mass significantly below the baryon budget, roughly consistent with the $\lesssim 20\%$ baryon fraction deduced from the cooling flow solutions in section~\ref{s:MW emission}.

\begin{figure}
 \includegraphics{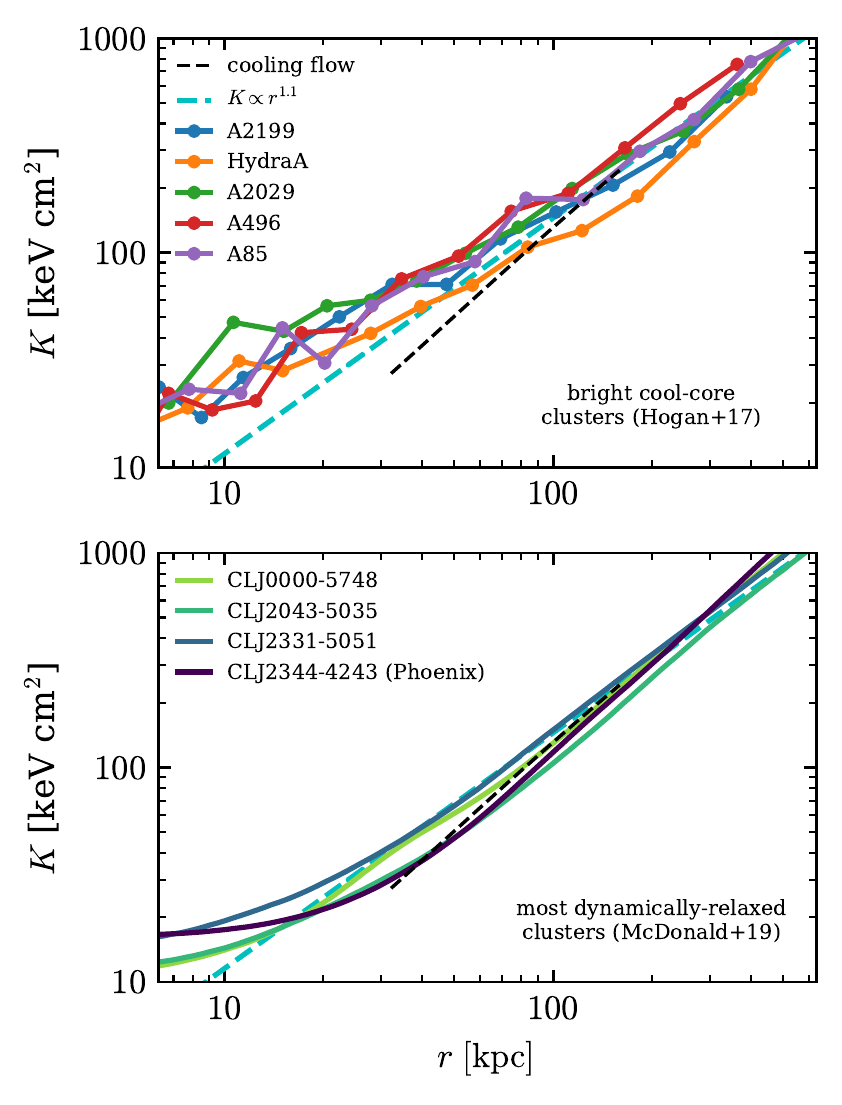}
\caption{Comparison of observed entropy profiles in cool-core clusters with a cooling flow solution. Top panel shows the deprojected entropy profiles of  five bright representative cool-core clusters from Hogan et al.~(2017), while the bottom panel shows model fits to the entropy profiles of the four most dynamically-relaxed clusters in the SPT sample (McDonald et al.~2019). 
Dashed cyan lines plot the scaling $K\propto r^{1.1}$ expected from self-similar cosmological accretion without cooling. The black lines plot a cooling flow solution spanning from $\Rhalf=30\kpc$ to $\Rcool=160\kpc$, where the normalization is chosen to roughly match the observations at $\Rcool$. 
In the top panel the observed profiles are flatter than expected in cooling flows. 
In contrast, the entropy profiles of Phoenix and CLJ2043-5035 in the bottom panel are apparently consistent with the cooling flow solution. 
}
\label{f:cluster observations}
\end{figure}

\subsection{Clusters}

The top panel in Figure~\ref{f:cluster observations} compares the predicted entropy profile in cooling flows with deprojected observations of five bright cool-core clusters from \cite{Hogan+17}. The dashed cyan line plots the scaling $K\propto r^{1.1}$ expected in self-similar cosmological accretion, i.e.\ when cooling is neglected (\citealt{TozziNorman01,Voit+05}). This prediction provides a good fit to the observations at large radii. The thick black line plots a cooling flow solution with $\Mhalo=10^{15}\msun$, $Z=0.3\zsun$, and $\Mdot=300\msun\yr^{-1}$, where the latter is chosen to match the observations at $\Rcool=160\kpc$. The plotted curve spans between $\Rhalf=30\kpc$ (eqn.~\ref{e:Rhalf}) and $\Rcool$. The entropy profile in the cooling flow scales as $r^{1.4}$ (eqn.~\ref{e:entropy cluster}), steeper than expected from gravity alone. 
Hence, without additional physical processes beyond gravity and cooling, the entropy profile is expected to steepen at $\Rcool$. In contrast, the observed entropy profile appears to flatten, reaching $K\propto r^{0.7}$ at small scales (\citealt{Panagoulia+14, Babyk+18}). 
The flattening of the profile appears to occur near $\Rcool$, suggesting that feedback affects the intracluster medium out to the maximum radius where it is actively cooling. 

The \cite{Hogan+17} clusters shown in the top panel were chosen mainly for their brightness, and are hence relatively representative of cool-core clusters. The bottom panel shows model fits to four objects from \cite{McDonald+19a}, which were selected to be the most dynamically-relaxed clusters out of the 100 clusters in the South Pole Telescope sample (SPT, \citealt{Bleem+15}). 
The entropy profiles at $30-160\kpc$ of these clusters is steeper than in the sample shown in the top panel, and similar to the slope predicted by the cooling flow solution, especially in the Phoenix and SPT-CLJ2043-5035 clusters. The intracluster medium (ICM) in the most dynamically relaxed clusters may hence be forming a cooling flow at these radii, i.e.~it may not be subject to significant heating by feedback. The entropy profiles flatten relative to the cooling flow solution within $\sim30\kpc \sim \Rhalf$.

In recent work published after the submission of this manuscript, \cite{McDonald+19b} compared a cooling flow solution directly to the observational data for the Phoenix cluster, rather than to a model fit of the data as done here. They found that a cooling flow provides a good match to the observations, supporting our conclusion that the ICM of this cluster may form a cooling flow.

\section{Discussion}\label{s:discussion}

In the previous sections, we demonstrated that initially hydrostatic gaseous halos converge onto one of a single-parameter family of cooling flow solutions, within a cooling time. In these solutions the volume-filling gas phase flows inward at a velocity $v(r)\approx -r/\tcool$, while the temperature is roughly equal to the circular temperature $\Tc\equiv\mu \mp \vc^2 / (\gamma\kb)$. These solutions are similar to the solutions developed for the intracluster medium in the 1980's (e.g.~\citealt{Fabian+84}). 

In Figure~\ref{f:cluster observations} we demonstrated that cooling flow solutions fail to explain the properties of the X-ray emission in typical clusters, as expected, though they are potentially consistent with the most dynamically-relaxed systems. 
On the other hand, we showed that a cooling flow solution 
with $\Mdot\sim{\rm SFR}$ correctly predicts the X-ray absorption line properties of the Milky Way halo (Fig.~\ref{f:MW absorption}), and the slope of the hot gas density profile derived by modelling the X-ray line emission in the MW halo (Fig.~\ref{f:MW density}).  Given that the cooling flow model is the simplest possible model and has no free parameters beyond the uncertainty in gas metallicity, this success may suggest that hot gas in the Milky Way halo forms a cooling flow with $\Mdot\sim{\rm SFR}$, at least in the inner 10s of kpc which dominate the observed X-ray emission and absorption. However, Figure~\ref{f:OVI} shows that similar solutions 
underpredict the observed \ion{O}{VI} by a factor of $\sim5$, both in the MW halo and in the halos of external star-forming galaxies. The observations of \ion{O}{VI}\ thus suggest that the cooling flow picture is incomplete. 

In this section we compare cooling flow solutions with other existing models of gas in galaxy-scale halos, and discuss possible resolutions to the apparently inconsistent conclusions implied by the X-ray and \ion{O}{VI} observations. 

\subsection{Comparison with hydrostatic models}

Several authors have proposed hydrostatic models for gas in galaxy-scale dark matter halos (\citealt{MallerBullock04, Faerman+17, MathewsProchaska17,Voit18}). Since the hydrostatic equation on its own is insufficient to fully specify the structure of the halo gas, these studies evoked additional assumptions on the density or entropy profile. Specifically, \cite{MallerBullock04} assumed a flat entropy profile out to $\Rcool$,
\cite{Faerman+17} derived the density profile from the average X-ray emission and absorption in the MW and from the \ion{O}{VI} absorption around external galaxies, \cite{MathewsProchaska17} assumed a density which follows the dark matter density with a flat entropy core, and \cite{Voit18} assumed an entropy profile which yields $\tcool/\tff\approx10$ at all radii. 
In the cooling flow solutions for the MW halo gas (Figs.~\ref{f:MW absorption} -- \ref{f:OVI}), the pressure profile is also close to hydrostatic, since the correction to the momentum equation is of order $\mach^2$, while $\mach$ is small at halo radii ($\approx 0.1$ at $100\kpc$, eqn.~\ref{e:mach Lstar}). However, although that the flow is relatively slow, it is this weak flow which sets the entropy profile of the halo gas via the entropy equation~(\ref{e:energy1}), at all radii smaller than $\Rcool$. Specifically, for Milky Way-mass halos we get $K\propto r^{0.87}$ (eqn.~\ref{e:self similar K} with $m=-0.1$).  Thus, in the absence of physical processes other than cooling, the entropy and density profiles in a quasi-static halo are not free parameters but rather set by the cooling flow solution.

\subsection{Reconciling OVI and X-ray observations}\label{s:discrepancy}

\begin{figure}
 \includegraphics{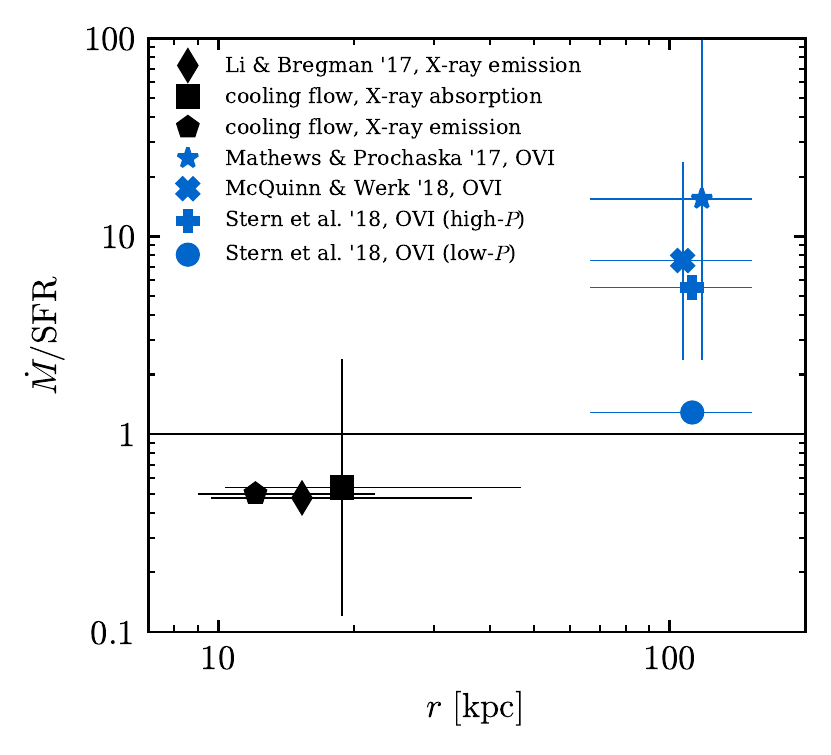}
\caption{Summary of gas cooling rates in low redshift $\sim L^*$ halos derived in this work and in previous studies, normalized by the SFR of the central galaxy. Black symbols mark values based on Milky-Way \ion{O}{VII} and \ion{O}{VIII} observations, while blue symbols mark values based on \ion{O}{VI} columns around external SF galaxies.  
The cooling flow-based estimates for the X-ray gas (black square and pentagon) are consistent with the estimate based on the phenomenological model of LB17 (black diamond), but are roughly an order of magnitude lower than the cooling rates implied by \ion{O}{VI} if it traces radiatively cooling gas (blue star, blue plus, and blue cross). This difference may indicate preferential heating of the \ion{O}{VI}-bearing outer halo. 
Alternatively, if \ion{O}{VI} traces thermal-equilibrium gas outside the accretion shock, the implied preshock mass flow rate (blue circle) is comparable to the postshock cooling rates suggested by the X-ray observations. 
}
\label{f:Mdot_to_SFR}
\end{figure}

Figure~\ref{f:Mdot_to_SFR} summarizes the constraints on $\Mdot$ derived in this study and in previous studies based on X-ray observations of \ion{O}{VII} and \ion{O}{VIII} (black), and based on UV observations of \ion{O}{VI} (blue). The horizontal position of each marker denotes the median distance contributing to the observations, while the range spans the $16-84$ percentiles. 
For the X-ray-based constraints this characteristic distance is estimated from the modelling, which for a density profile of $\nH\propto r^{-1.6}$ predicted by cooling flows (eqn.~\ref{e:nH Lstar}) yields an X-ray gas distance of $\sim10-40\kpc$. 
As discussed in section \ref{s:MW absorption}, comparison of the cooling flow solution with X-ray absorption yields $\Mdot\approx 0.12-2.4$ times the MW SFR of $1.6\msun\yr^{-1}$ (black square in Fig.~\ref{f:Mdot_to_SFR}), while a comparison with the X-ray line emission yields a cooling rate of roughly half the SFR (black pentagon, see section \ref{s:MW emission} and Fig.~\ref{f:MW density}).
These deduced density profile, characteristic distance, and cooling rate of the X-ray gas are all consistent with the results of LB17 (black diamond), who modelled the X-ray observations using a phenomenological model independent of our cooling flow solutions. The results of LB17 are based on and consistent with other results by the same group (\citealt{MillerBregman13,MillerBregman15,Bregman+18,QuBregman18}). The cooling flow solution hence provides a physical explanation for the parameters implied by LB17's phenomenological analysis.

In contrast with the relative proximity implied for the X-ray gas, a deprojection of \ion{O}{VI}\ observations around external galaxies shows that \ion{O}{VI}\ absorption originates in gas farther out in the halo, at a radius range of $0.35-0.8\Rvir$, or $70-150\kpc$ for the median $\Rvir$ of $190\kpc$ in the COS-Halos sample (fig.~1 in S18, see consistent result in \citealt{MathewsProchaska17}). 
Cooling flow solutions with $\Mdot={\rm SFR}$ underpredict the observed \ion{O}{VI} absorption by a factor of $\sim 5$, and there is no stable cooling flow solution which can reproduce the observed \ion{O}{VI} (Fig.~\ref{f:OVI}, section~\ref{s:OVI}). Fig.~\ref{f:Mdot_to_SFR} plots previous estimates of the \ion{O}{VI}-traced $\Mdot$ assuming that it originates in radiatively-cooling gas, from \citeauthor{MathewsProchaska17} (2017, blue star), from \citeauthor{McQuinnWerk18} (2018, blue cross) and from S18 (their `high-pressure' scenario, blue plus). In all cases, we normalized the mass inflow rates found by these studies by the mean ${\rm SFR}=4.2\msun\yr^{-1}$ in the observed sample (\citealt{Werk+13}). The derived $\Mdot/{\rm SFR}$ are roughly an order of magnitude larger than implied by the X-ray emission. 

The \ion{O}{VI} columns and dispersion observed around the MW are similar to those around external SF galaxies, if one accounts for the different perspectives and for confusion with the Galactic disk (\citealt{Zheng+15}, Fig.~\ref{f:OVI}). It thus seems likely that \ion{O}{VI} in the Milky Way has the same origin as \ion{O}{VI} in other SF galaxies. Fig.~\ref{f:Mdot_to_SFR} therefore suggests a somewhat surprising result that cooling rates in the \ion{O}{VI}-bearing outer halo are significantly different from cooling rates in the inner halo which dominates the X-ray observations. It is important to note that this conclusion is already implied by the comparison of the LB17 and e.g.~\cite{McQuinnWerk18} data points even without considering the cooling flow analysis in this work, as discussed in \cite{QuBregman18}. Our cooling flow analysis strengthens this conclusion by providing a physical basis for LB17's phenomenological results, as mentioned above. 
We next discuss several possible resolutions to this apparent tension between the X-ray and \ion{O}{VI} observations.

{\bf Preferential heating of the OVI-bearing outer halo:}
To maintain the large cooling rates implied by the \ion{O}{VI} observations, previous studies typically assumed that gas at $\gtrsim100\kpc$ is heated by supernovae or AGN in the central galaxy (\citealt{Cen13,Liang+16,MathewsProchaska17,Faerman+17,Suresh+17,McQuinnWerk18}). 
If this is the case, a heating mechanism powered by the central galaxy must avoid depositing most of its energy in the inner 10s of kpc to explain the order-of-magnitude lower cooling rates inferred from the X-ray observations. 
Alternatively, the heating energy may originate from larger radii, e.g.\ processes associated with cosmological structure assembly. 

In the middle panel of Fig.~\ref{f:OVI} we plot a potential \ion{O}{VI} profile in the presence of a heating mechanism that preferentially heats the outer halo (black dash-dotted line). To produce this profile we run a simulation similar to the simulations shown in Fig.~\ref{f:snapshots}, with $\Mhalo=6\cdot10^{11}\msun$ and an initial baryon mass of $0.4$ times the halo baryon budget. At each assumed impact parameter, we integrate the \ion{O}{VI} volume density in the $t=7\Gyr$ snapshot out to a maximum radius of $200\kpc$. This snapshot is chosen since it has a relatively large cooling rate of $20\msun\yr^{-1}$ within $200\kpc$, compared to a lower cooling rate of $4\msun\yr^{-1}$ within $40\kpc$, thus mimicking the different $\Mdot$ implied by the \ion{O}{VI} and X-ray observations shown in Fig.~\ref{f:Mdot_to_SFR}. Figure~\ref{f:OVI} demonstrates that in this snapshot the \ion{O}{VI} columns are roughly consistent with observations. In our simulations which do not include heating mechanisms, such a gas structure is necessarily a transient phenomenon, which occurs only when $\d\Rcool/\d t$ increases quickly with time and there is hence a disparity between $\Mdot$ at large scales and $\Mdot$ at small scales (see eqn.~\ref{e:Mdot(t)}). A transient solution is unlikely to explain the observed \ion{O}{VI} which are seen around practically all blue galaxies. Potentially, in the presence of a heating mechanism which preferentially heats the outer halo as suggested here, such an \ion{O}{VI} profile may be long-lived and thus consistent with the observations.

{\bf The low-pressure OVI scenario:} The assumption that \ion{O}{VI}\ traces radiatively cooling gas at $T\approx10^{5.5}\K$
is mainly based on the results of cosmological simulations (\citealt{Stinson+12, Hummels+13, Cen13,Oppenheimer+16, Liang+16, Gutcke+17,Suresh+17,Nelson+18}). 
From a purely observational perspective, \ion{O}{VI} along random quasar sight lines and in sight lines farther than $0.5\Rvir\approx100\kpc$ from galaxies is almost always observed together with \ion{H}{I}, with a column ratio of $N_{\rm HI}/\Novi\approx 1-10$. These \ion{H}{I} observations support a scenario where the \ion{O}{VI} and associated \ion{H}{I} columns trace single-phase photoionized gas in thermal equilibrium with the UV background (\citealt{Tripp+08, ThomChen08, Stern+16}, S18). If the \ion{O}{VI}-gas is in thermal equilibrium, then it is not radiatively cooling, and hence does not imply cooling rates significantly larger than suggested by the X-ray observations. 

\ion{O}{VI} can trace gas in thermal equilibrium if the gas has a relatively low thermal pressure of $\lesssim1\cm^{-3}\K$, as expected outside the accretion shock. S18 argued that this would require a shock radius around $\sim$$L^{\star}$ galaxies $\Rshock\lesssim 0.5 R_{\rm vir}\approx100\kpc$, substantially smaller than predicted by cosmological simulations. They showed that such a scenario is consistent with the observed \ion{C}{III} absorption and lack of low-ion absorption at $R_{\perp}\gtrsim0.5\Rvir$, and with the observed linear relation between \ion{O}{VI} column and velocity width. 
Also, if \ion{O}{VI} traces free-falling preshock gas then the implied mass inflow rate is $\approx 5\msun\yr^{-1}$ (eqn.~30 in S18), comparable to the mean ${\rm SFR}=4.2\msun\yr^{-1}$ of the central galaxies. This alternative estimate of $\Mdot/{\rm SFR}\approx1.2$ implied by \ion{O}{VI} is plotted in Fig.~\ref{f:Mdot_to_SFR} (blue circle), and is comparable to the $\Mdot/{\rm SFR}\sim0.5$ suggested by the MW X-ray observations, thus alleviating the tension implied by assuming that \ion{O}{VI} traces radiatively-cooling gas. 

In the middle panel of Figure \ref{f:OVI} we plot a possible $\Novi$ profile based on the model suggested by S18 (black dashed line), calculated as follows. 
For the hot gas inside the shock radius, we calculate a cooling flow solution with the mean $\Mhalo=6\cdot10^{11}\msun$ and $z=0.2$ of the galaxies in the panel, $\Mdot$ equal to their average SFR of $3\msun\yr^{-1}$, and an  assumed metallicity of $0.5\zsun$. We require the outer boundary of the cooling flow to satisfy shock jump conditions with zero shock velocity. There is only a single possible solution for jump conditions consistent with preshock gas in thermal equilibrium with the UV background ($\approx 3\cdot10^4\K$). This solution is plotted in Figure~\ref{f:t_ratio_shock} in the appendix. 
For the gas outside the shock radius, the jump conditions imply the same $\Mdot=3\msun\yr^{-1}$ as in the postshock gas and a preshock velocity\footnote{The characteristic velocity offset between the central galaxies and the \ion{O}{VI} absorption profiles is $\approx100\kms$ (\citealt{Tumlinson+11}), lower than the deduced radial velocity of $150\kms$ for the \ion{O}{VI} gas in this model. This difference may potentially be explained by projection effects. Alternatively, assuming a preshock gas temperature somewhat higher than equilibrium with the background (e.g.~due to adiabatic compression), would imply a lower preshock velocity which is more consistent with the observations.} of $150\kms$.  
We assume a velocity profile of $v\propto r^{-1/2}$ and a clumping factor of $\delta\rho/\rho=4$, which yield a best `by-eye' fit to the \ion{O}{VI} observations.
This clumping factor is consistent with the \ion{O}{VI} absorber pathlengths of $\gtrsim10$s of kpc implied by observations relative to the total pathlength through the outer halo (Table~1 in S18). 
As can be seen in Figure~\ref{f:OVI}, the expected $\Novi$ in this combined free-fall + cooling flow model is roughly consistent with the observed $\Novi$, while also matching the $\Mdot\sim{\rm SFR}$ suggested by the MW X-ray observations. Predictions of this model for observations of \ion{Ne}{VIII} and other ions observable in the extreme UV (EUV) are discussed in \cite{Stern+18}.

{\bf A density profile shallower than a cooling flow:} 
An alternative resolution 
is that the halo density profile is shallower than deduced by LB17 from the MW observations (see section 2.4 in \citealt{Bregman+18} for a discussion of this possibility). A flatter density profile would yield a larger characteristic radius and higher cooling rates for the X-ray gas. Such a flat density profile was deduced by \cite{Faerman+17}, who fit a hydrostatic model to the average of the X-ray observations in the MW and to the \ion{O}{VI} observations around external galaxies. The density and mass profile in their model are plotted in Figure~\ref{f:MW density}. 
Note that \citeauthor{Faerman+17}\ did not incorporate the dependence of X-ray observations on direction in the halo, which was used by LB17 to infer the slope of the density profile.
In the \cite{Faerman+17} model, both the X-ray and \ion{O}{VI} trace cooling rates of $\approx30\msun\yr^{-1}$. This model suggests both significant feedback heating in the halo and a closed baryon fraction (see lower panel of Figure~\ref{f:MW density}), in contrast with the baryon-deficient halo and lack of heating implied by the cooling flow model and by LB17.

\subsection{Comparison with precipitation models}\label{s:precipitation}

We demonstrate in Figure~\ref{f:sigma rho} and section~\ref{s:sims} that small amplitude perturbations in a cooling flow do not develop into multi-phase structure at radii where $\tcool/\tff>1$, consistent with previous results based on linear perturbation theory (e.g.~\citealt{BalbusSoker89}) and hydrodynamic simulations (e.g.~\citealt{Joung+12, Sharma+12a}). 
In this regime, the gas is predominantly single phase (by volume) and cool clouds develop only near strong disturbances which can seed non-linear perturbations (see section~\ref{s:cool clouds}). 

However, numerical simulations have demonstrated that in the presence of heating by feedback multi-phase structure may develop even if $\tcool/\tff$ is as high as $\sim10$ (\citealt{Sharma+12a, Gaspari+13, LiBryan14b, ChoudhurySharma16,Choudhury+19}). Based on this result \cite{Sharma+12b} and \cite{Voit+15, Voit+17} proposed a feedback-regulated limit cycle. 
In this cycle, cool clumps in the halo `precipitate' onto the central galaxy, providing fuel for feedback which heats the ambient medium and promotes further condensation. The main ansatz of these precipitation models, based on the results of numerical simulations, is that the feedback loop regulates the halo gas to have a minimum $\tcool/\tff\sim10$. Some support for this ansatz is provided by observations of X-ray emitting gas in groups and clusters of galaxies, which apparently adhere to the limit of $\tcool/\tff\gtrsim10$ and exhibit evidence for multiphase gas preferentially in objects and at radii where $\tcool/\tff$ is close to the limit (\citealt{McCourt+12,VoitDonahue15}). 

In galaxy-scale halos, the roughly constant $\tcool/\tff$ suggested by precipitation models is expected also in the absence of any feedback (eqn.~\ref{e:tcool to tff Lstar} and lower-right panel of Figure~\ref{f:by density}). This result follows from the cooling flow solution (eqns.~\ref{e:self similar T} -- \ref{e:self similar nH}), which yields $\tcool/\tff\propto r^{0.5+2m}$. For the weakly decreasing circular velocity profile expected in galaxy-scale halos ($\vc\propto r^{m}\propto r^{-0.1}$, see Figure~\ref{f:vc and Lambda}), we get $\tcool/\tff$ which depends only weakly on radius. Thus, a roughly constant $\tcool/\tff$ in simulations or implied by observations does not necessarily imply that galaxy-scale halos are precipitation-regulated. 

In cooling flow solutions, the normalization of $\tcool/\tff$ is set by the free parameter of the solution and in principle can have any value larger than unity. In practice, we expect the total hot gas mass -- and thus the normalization of $\tcool/\tff$ -- to be set either by the cosmic baryon budget or modified by the loss of baryons driven by galactic outflows at high redshift. 
Interestingly, for a Milky Way-mass halo, assuming $\Mdot=1\msun\yr^{-1}\approx{\rm SFR}$ implies a characteristic $\tcool/\tff=7.5$ (eqn.~\ref{e:tcool to tff Lstar}), approximately equal to the basic ansatz of precipitation models. 
Thus precipitation-regulated models for the MW yield roughly the same halo gas structure as the no-ongoing-feedback cooling flow solution discussed in this work. This similarity can be seen in the top panel of Figure~\ref{f:MW density}, where the density profile of a precipitation-regulated halo (\citealt{Voit18}, black dash-dotted line) is compared to the density profile of the cooling flow solution (cyan line). The two profiles differ by less than a factor of two at all plotted radii, within the uncertainty of the modelling and of the observational constraints. Thus, the predictions of precipitation-regulated models for the hot gas structure and their comparison with observations (e.g.~\citealt{Voit18,Voit+19}) do not differentiate between a precipitation-regulated CGM and a cooling flow CGM. 

\subsection{Additional implications for feedback}\label{s:time dependence}

Figures~\ref{f:MW absorption}--\ref{f:MW density} demonstrate that observational constraints on the hot gas in the MW halo are consistent with the predictions of a cooling flow with $\Mdot\approx{\rm SFR}$. In this scenario ongoing heating by feedback is weak, and does not significantly change the halo gas properties. Strong feedback is however still required in this scenario at earlier epochs to explain the high metallicity ($0.3\zsun-\zsun$) and low gas mass ($\lesssim20\%$ of the cosmic baryon budget, section~\ref{s:MW emission}) of the cooling flow solutions. Such a scenario where feedback is strong at high redshift but weak in the local Universe is qualitatively consistent with observational indications that galactic winds are strong in high SFR surface density galaxies common at $z\sim2$, while they are weak in the low SFR surface density galaxies common at $z\sim0$ (e.g., \citealt{HeckmanThompson17}). This scenario is also (qualitatively) supported by cosmological simulations that predict strong winds at $z\sim2$ which subside by $z\sim0$ in $\sim L^*$ halos (e.g., in the FIRE cosmological simulations, \citealt{Muratov+15,Muratov+17}). 

Another requirement of the cooling flow model is that ongoing heating by the central black hole is inefficient. In the MW, possible evidence for such heating are the \textit{Fermi} bubbles (\citealt{Su+10}), which extend $\approx10\kpc$ from the Galactic center. 
\cite{Sarkar+17} estimated a feedback energy injection rate into the bubbles of $\approx(0.7-1)\cdot10^{41}\erg\s^{-1}$. 
If this energy eventually heats the halo gas, it could offset the energy radiated by a cooling flow with $\Mdot=1.5-2\msun\yr^{-1}$ at radii $10-100\kpc$. Hence for feedback heating to be small relative to $\Mdot\approx1.6\msun\yr^{-1}$, this heating rate estimate needs to be somewhat biased high, or alternatively the bubble energy is not efficiently transferred to the surrounding CGM.

In cluster-scale halos, the similarity of the entropy profiles of Phoenix and CLJ2043-5035 with cooling flow solutions may suggest that these clusters are in a cooling flow phase of a feedback limit cycle (see introduction), as also supported by the large UV-estimated SFRs in their central galaxy ($\approx2000$ and $\approx200\msun\yr^{-1}$, respectively, \citealt{McDonald+12, McDonald+16}). It would be useful to conduct a similar comparison of the cooling flow solutions with simulations of feedback limit cycles (e.g.~\citealt{Prasad+15}) and thus derive the predicted duty cycle of the cooling flow phase in such models. This prediction could then be compared to the observed fraction of clusters in the cooling flow phase. 

\section{Summary}\label{s:summary}

In this paper, we solve the spherical steady-state equations of radiatively cooling gas, in a time-independent gravitational potential characteristic of dark matter halos. Our solutions extend previous solutions derived in the 1980's for cluster-scale halos to galaxy-scale halos. We derive self-similar solutions in the hydrostatic limit for power-law potential and cooling function, and numerical solutions for more general potentials and cooling functions. 
Both type of solutions have a single free parameter, and are thus fully determined if either the mass inflow rate or total halo gas mass are known. We find that in cooling flows the entropy scales with radius as $K\propto r^{1+4m/3}$ (eqn.~\ref{e:dlnKdlnr}), where $m\equiv\d\log\vc/\d\log r$ is the power-law index of the circular velocity radial profile. 

Using idealized 3D hydrodynamic simulations of initially hydrostatic gas, we showed that the average gas properties in the simulation converge onto a cooling flow solution within a cooling time or, equivalently, out to the cooling radius (Figs.~\ref{f:M15 sims} -- \ref{f:M12 sims high density}). Density fluctuations remain significantly below unity as long as $\tcool>\tff$ (Fig.~\ref{f:sigma rho}), consistent with expectations based on linear theory. Halo gas which forms a cooling flow is thus expected to be predominantly single phase, with cool clouds limited to strong disturbances that can seed non-linear perturbations (e.g.\ near the disk plane or due to stirring by satellite galaxies). 

We compare the cooling flow solutions with observational constraints on halo gas at low redshift. Our conclusions can be summarized as follows:
\begin{enumerate}

 \item Observations of \ion{O}{VII} and \ion{O}{VIII} absorption in the Milky Way halo are consistent with a cooling flow solution with $\Mdot\sim{\rm SFR}$, with no free parameters (Fig.~\ref{f:MW absorption}). 
 
 \item The Milky Way halo gas density profile of $n_{\rm e}\propto r^{-1.5}$, deduced by \cite{LiBregman17} from modelling of observed \ion{O}{VII} and \ion{O}{VIII} emission, is consistent with the prediction of cooling flows (Fig.~\ref{f:MW density}). 
 
 \item Estimates of the thermal pressure based on cool clouds in the ISM and in the Magellanic Stream, and the dispersion measure towards the LMC, are also consistent with a cooling flow solution with $\Mdot\sim{\rm SFR}$.
  
 \item Cooling flows with $\Mdot\sim{\rm SFR}$ underpredict observed \ion{O}{VI} absorption columns in the MW halo and in the halos of other star-forming galaxies, typically by a factor of $\sim 5$, if the cooling flows are assumed to extend out to $\gtrsim \Rvir$  (Fig.~\ref{f:OVI}). This discrepancy can be reconciled with the successes of the cooling flow solution described in (i)--(iii) if the \ion{O}{VI}-bearing outer halo is preferentially heated or, alternatively, if the large $\Novi$ originate in free-falling gas outside an accretion shock at $\lesssim100\kpc$ which subsequently develops into a cooling flow at smaller radii, as suggested by \cite{Stern+18}. Upcoming observations of \ion{Ne}{VIII} and other ions observable in the EUV could potentially test these possibilities.

\item The cooling flow solution for the MW halo gas predicts $\tcool/\tff\approx 7.5 (r/100\kpc)^{0.3}$, similar to the basic ansatz of thermal-instability-regulated feedback-loop models, in which $\tcool/\tff\approx10$ independent of radius.
 These feedback-loop models thus yield roughly the same hot gas structure and observables as the no-feedback cooling flow solution (Fig.~\ref{f:MW density}). 
 
 \item In galaxy cluster halos, cooling flows predict a radial entropy profile of $K\propto r^{1.4}$, steeper than the observed profile of $K\propto r^{0.7}$ in typical cool-core clusters. This is consistent with other evidence that feedback alters the gas properties in the intracluster medium. However, two of the most dynamically-relaxed clusters in the 
 SPT sample have entropy profiles consistent with cooling flows beyond the galaxy scale $\approx30\kpc$ (Fig.~\ref{f:cluster observations}). These clusters may be in the cooling flow phase of a feedback limit cycle. 
 \end{enumerate}

The possibility that hot gas in Milky-Way-like halos forms a cooling flow with negligible ongoing feedback heating, could be a consequence of the fact that galactic winds are weak in galaxies with low SFR surface density (e.g., \citealt{HeckmanThompson17}). This is also suggested by simulations that predict weak winds in low-redshift $\sim$$L^\star$ galaxies (e.g., \citealt{Muratov+15}), as discussed in section~\ref{s:time dependence}. 
Our analysis thus raises the question of which halos (as a function of mass and redshift) can be adequately modeled by a cooling flow over an interesting range of radii. 
At high redshift, stellar feedback is expected to be stronger and is also required to explain the high metallicity and low gas mass implied by the cooling flow solutions at low redshift. 
Comparison of cooling flow solutions with observations of high-redshift halos (e.g.~\citealt{Turner+14,Rudie+19}) may be able to address this important question.

\section*{Acknowledgements}

We thank the referee, Prateek Sharma, for a thorough report and illuminating comments that significantly improved the paper. We thank also Mark Voit, Sean D.~Johnson,  and Zachary Hafen for detailed and insightful comments. We thank Michael McDonald for the entropy profiles fits used in Fig.~\ref{f:cluster observations}. 
JS is supported by the CIERA Postdoctoral Fellowship Program. DF is supported by the Flatiron Institute, which is supported by the Simons Foundation. CAFG is supported by NSF through grants AST-1517491, AST-1715216, and CAREER award AST-1652522, by NASA through grants NNX15AB22G and 17-ATP17-0067, by STScI through grants HST-GO-14681.011, HST-GO-14268.022-A, and HST-AR-14293.001-A, and by a Cottrell Scholar Award from the Research Corporation for Science Advancement. This work was supported in part by a Simons Investigator Award from the Simons Foundation and by NSF grant AST-1715070.


\appendix

\section{Integration details}\label{a:shooting}

\begin{figure*}
 \includegraphics{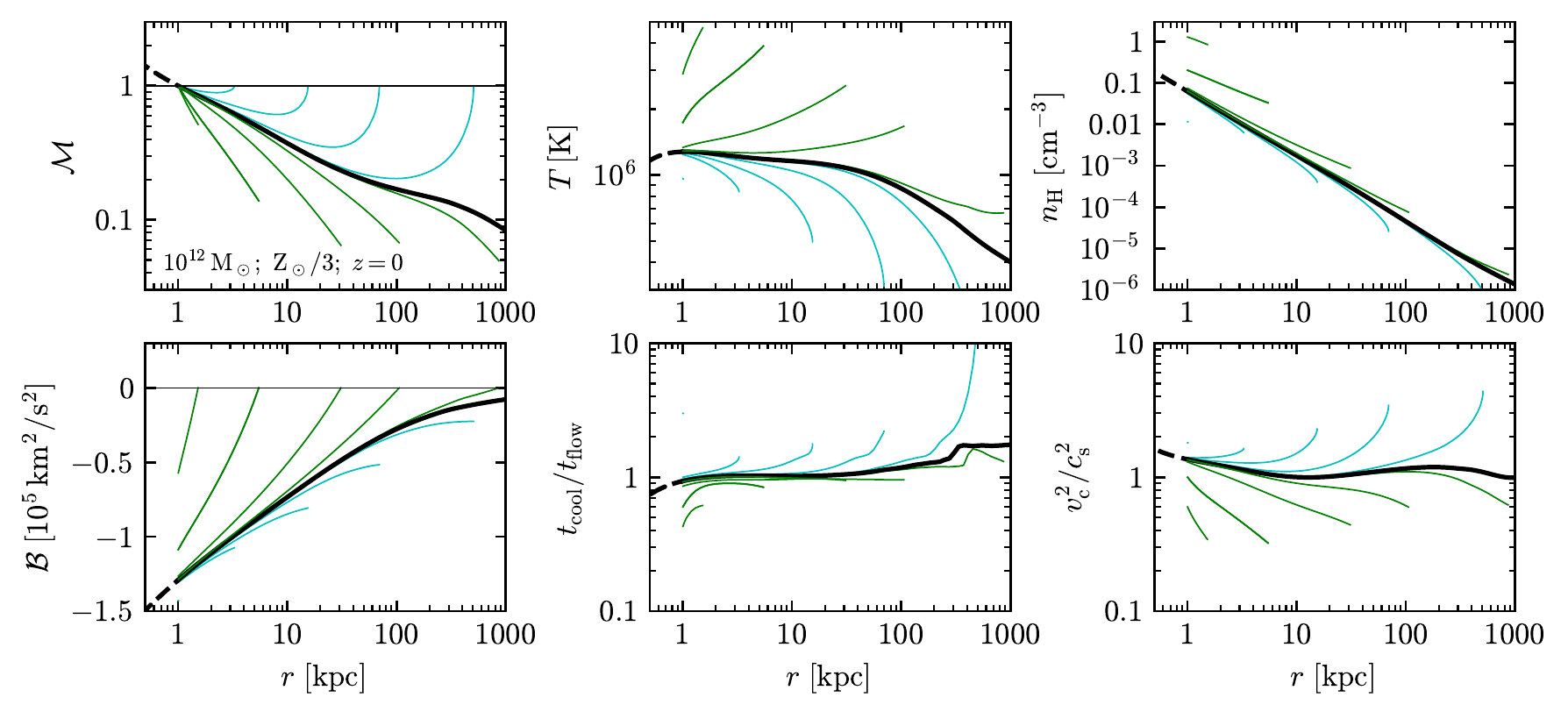}
\caption{An example of the shooting method used in this work. 
The panels show the Mach number, temperature, density, Bernoulli parameter, cooling time to flow time ratio, and ratio of the circular velocity to the sound speed squared. Different lines show different integrations of the steady-state flow equations, assuming third-solar metallicity gas in a $10^{12}\msun$ halo at $z=0$.
Integration starts at the sonic point, assumed to be at $\Rsonic=1\kpc$, and proceeds outwards. 
The temperature at $\Rsonic$ is constrained via the shooting method. 
If the solution becomes unbound ($\Bernoulli>0$, green curves) then $T(\Rsonic)$ is adjusted downward. 
If the solution reaches $\mach=1$ with an infinite velocity derivative (cyan curves) then $T(\Rsonic)$ is adjusted upward. Only for a narrow range of $T(\Rsonic)$ the integration reaches large radii with $\Bernoulli<0$ (black curves). These `marginally-bound' solutions form the single-parameter family of solutions used in this work. 
}
\label{f:integration example}
\end{figure*}

In this section we describe the numerical integration technique used to derive the marginally-bound transonic solutions.

We first integrate the flow equations outward from a sonic point at an assumed radius $\Rsonic$, in order to derive the subsonic part of the solution. In the integration we use the form of the flow equations expressed in eqns.~(\ref{e:mass1}), (\ref{e:dlnKdlnr}), and (\ref{e:dlnvdlnr}).
The temperature at $\Rsonic$ is initially assumed to equal $\Tc$, while $\rho(\Rsonic)$ and $v(\Rsonic)$ are derived from this assumption and from the requirement that both sides of eqn.~(\ref{e:dlnvdlnr}) vanish. Since the momentum equation is indeterminate at $\Rsonic$, we start the integration at a radius $R_0=(1+\epsilon)\Rsonic$ with $\epsilon=10^{-5}$. The values of the hydrodynamic variables $T(R_0)$, $v(R_0)$, and $\rho(R_0)$ thus differ from their values at $\Rsonic$ by factors of $1+(\d\ln T/\d\ln r)\epsilon$, $1+(\d\ln v/\d\ln r)\epsilon$, and $1-(\d\ln v/\d\ln r+2)\epsilon$, respectively. The derivation of $\d\ln T/\d\ln r$ and $\d\ln v/\d\ln r$ for a given $T(\Rsonic)$ is described below. 
If the integrated solution reaches $\mach=1$ with an infinite velocity derivative, then $T(\Rsonic)$ is adjusted upward, and the flow equations are integrated again. If the solution becomes unbound ($\Bernoulli>0$), then $T(\Rsonic)$ is adjusted downward and again the integration is repeated. The process continues until we find a solution which is bound out to a large radius of $10\Rvir$. Then, using $T(\Rsonic)$ of this marginally-bound solution we find also the supersonic part of the solution, by applying a similar offset to the hydrodynamic variables with $\epsilon=-10^{-5}$, and integrating inward.  
An example of this process is shown in Figure~\ref{f:integration example}. The derived marginally-bound solution in this example is the solution shown in Fig.~\ref{f:example}. 

To derive $\d\ln T/\d\ln r$ and $\d\ln v/\d\ln r$ at the sonic point we divide eqn.~(\ref{e:dlnvdlnr}) by $\mach^2-1$, which gives
\begin{equation}\label{e:vtag}
 \vtag = \frac{2\cs^2-\vc^2- \frac{\cs^2\tflow}{\gamma\tcool}}{v^2-\cs^2} ~.
\end{equation}
For the velocity derivative to be finite at the sonic radius (where $v=-\cs$), the numerator must equal zero, i.e.
\begin{equation}\label{e:x def}
 \frac{\tflow}{\gamma\tcool} = 2 - \frac{\vc^2}{\cs^2} \equiv 2(1-x) ~,
\end{equation}
where for convenience we defined the parameter $x\equiv \vc^2/2\cs^2=\Tc/2T$. Note that for adiabatic conditions ($\tflow/\tcool\rightarrow0$),  condition~(\ref{e:x def}) is equal to the standard Bondi condition of $\vc^2=2\cs^2$ or $x=1$. The logarithmic derivative of $v$ at the sonic point is then found using l'Hospital's rule: 
\begin{equation}\label{e:vtag st}
  \left(\frac{\d \ln v}{\d \ln r}\right)_{\rm sonic\,point} = \frac{\frac{\rm d}{\rm d r}\left(2\cs^2-\vc^2- \frac{\cs^2\tflow}{\gamma\tcool}\right)}{\frac{\rm d}{\rm d r}\left(v^2-\cs^2\right)} ~.
\end{equation}
The denominator is equal to 
\begin{equation}\label{e:denominator derivative0}
 \frac{\rm d}{\rm d r}\left(v^2-\cs^2\right) = \frac{2v^2 \vtag}{r} - \frac{\cs^2 \Ttag}{r} = \frac{\cs^2}{r}\left(2\vtag-\Ttag\right) ~,
\end{equation}
where in the second equality we used $-v=\cs$. The relation between the logarithmic derivative of $v$ and the logarithmic derivative of $T$ can be derived from mass and entropy conservation (eqns.~\ref{e:dlnrho} and \ref{e:dlnKdlnr}):
\begin{equation}\label{e:Ttag}
 \Ttag + (\gamma-1)\left(2+\vtag\right) = \frac{\tflow}{\tcool} 
\end{equation}
Using the sonic point equality (\ref{e:x def}) for $\tflow/\tcool$, assuming $\gamma=5/3$ and rearranging we get
\begin{equation}\label{e:vtag from Ttag}
 \vtag = -\frac{3}{2}\Ttag + 3 - 5x ~.
\end{equation}
Plugging this equation in eqn.~(\ref{e:denominator derivative0}) then yields for the denominator of eqn.~(\ref{e:vtag st})
\begin{equation}\label{e:numerator derivative}
 \frac{\rm d}{\rm d r}\left(v^2-\cs^2\right) = \frac{\cs^2}{r}\left[-4\Ttag +6-10x\right] ~.
\end{equation}
For the numerator in eqn.~(\ref{e:vtag st}) we note that the term $\cs^2\tflow/\gamma\tcool$ is proportional to $r\rho\Lambda/-v$. Hence for $\vc=\vc(r)$ and $\Lambda=\Lambda(T,\rho)$ we get
\begin{eqnarray}
 & \frac{\rm d}{\rm d r}\left(2\cs^2-\vc^2 - \frac{\cs^2\tflow}{\gamma\tcool}\right) = \frac{2\cs^2\Ttag}{r} - \frac{2\vc^2\vctag}{r} 
 \nonumber\\
 & -\frac{1}{r}\frac{\cs^2\tflow}{\gamma\tcool}\left[1+\rhotag\left(1+\Lambdatagrho\right)+\Ttag\Lambdatag - \vtag\right]  ~.
\end{eqnarray}
Using again mass conservation ($\d\ln\rho/\d\ln r = -\d\ln v/\d\ln r-2$) and the sonic point equality (\ref{e:x def}) we get 
\begin{eqnarray}\label{eq: vtag st1}
 &  \frac{\rm d}{\rm d r}\left(2\cs^2-\vc^2-\frac{\cs^2\tflow}{\gamma\tcool}\right) = \frac{\cs^2}{r}\left\{2\Ttag - 4x\vctag\right.
  \nonumber\\
    &\left.+2\left(1-x\right)\left[\vtag\left(2+\Lambdatagrho\right)+1+2\Lambdatagrho-\Ttag\Lambdatag\right]\right\} ~.
\end{eqnarray}
Using eqn.~(\ref{e:vtag from Ttag}) in eqn.~(\ref{eq: vtag st1}) and rearranging we get for the numerator of eqn.~(\ref{e:vtag st})
\begin{eqnarray}\label{e:denominator derivative}
& \frac{\rm d}{\rm d r}\left(2\cs^2-\vc^2- \frac{\cs^2\tflow}{\gamma\tcool}\right) = \nonumber\\
& \frac{\cs^2}{r} 
\left[\Ttag\left(2-6(1-x)-2(1-x)\left(\Lambdatag+\frac{3}{2}\Lambdatagrho\right)\right)\right.  \nonumber\\
 & \left.- 4x\vctag + 2(1-x)\left(7-10x+\left(5-5x\right)\Lambdatagrho\right)\right] ~.
\end{eqnarray}
Finally, using equations~(\ref{e:vtag from Ttag}), (\ref{e:numerator derivative}), and (\ref{e:denominator derivative}) in eqn.~(\ref{e:vtag st}) we get that
\begin{eqnarray}\label{e:quadratic}
& \left(\Ttag\right)^2 +\Ttag\left[\frac{29}{6}x-\frac{17}{6}+\frac{1}{3}(1-x)\left(\Lambdatag+\frac{3}{2}\Lambdatagrho\right)\right]\nonumber\\
& + \frac{2}{3}x\vctag + 5x^2-\frac{13}{3}x+\frac{2}{3}-\frac{5}{3}(1-x)^2\Lambdatagrho = 0 ~,
\end{eqnarray}
which can be solved as a quadratic equation for a given $x=\vc^2/2\cs^2$, $\d\ln\Lambda/\d\ln T$, $\d\ln\Lambda/\d\ln \rho$, and $\d\ln\vc/\d\ln r$. 
Typically, one of these roots has $\d\mach/\d r<0$, which corresponds to a cooling flow solution that is subsonic at large scales and supersonic at small scales, while the other root has $\d\mach/\d r>0$, which corresponds to the opposite transition. We use the former root in the solutions used in this paper. 

A similar derivation was done by \cite{MathewsGuo12} for the specific case of a point mass and constant cooling function, and their result can be reproduced using $\d\ln\Lambda/\d\ln T=\d\ln\Lambda/\d\ln\rho=0$ and $\d\ln\vc/\d\ln r=-1/2$ in eqn.~(\ref{e:quadratic}). Note that the parameter `a' defined by \citeauthor{MathewsGuo12} is equal to $(1-x)/x$ in our notation.

\section{Outer boundary conditions}\label{a:boundary}

\begin{figure*}
 \includegraphics{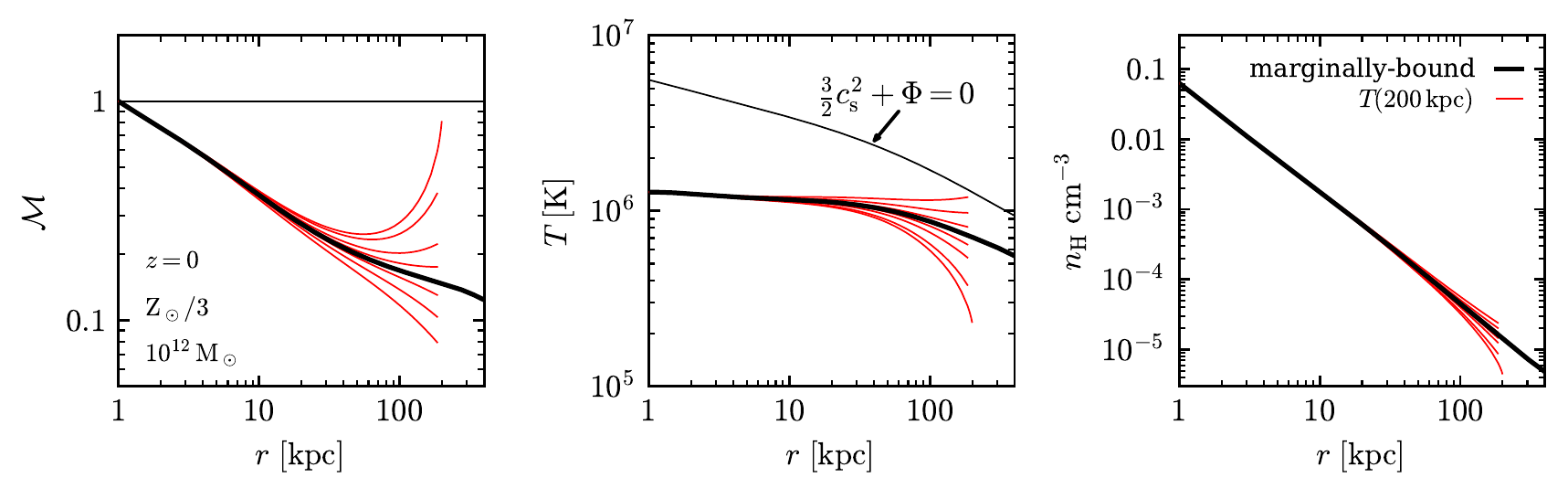}
\caption{Transonic solutions with the same $\Mdot$ and different choices of the boundary temperature at $200\kpc$. All assumed outer boundary conditions are subsonic and bound. The temperature of static gas with $\Bernoulli=0$ is marked in the middle panel. A marginally-bound solution as used in this work is plotted as a thick black line. This Figure demonstrates that the choice of the outer boundary condition significantly affects the solution only near the outer boundary. } 
\label{f:outer boundary}
\end{figure*}

The outer boundary condition of the transonic marginally-bound solutions used in this work is $\Bernoulli\rightarrow 0^{-}$ as $r\rightarrow\infty$ (see Appendix~\ref{a:shooting}). How will the solution change if we impose a different outer boundary condition?
In Figure~\ref{f:outer boundary} we repeat the integration process shown in Fig.~\ref{f:integration example}, but applying the shooting method until some specific temperature at $200\kpc$ is reached. The marginally-bound solution is marked by a thick black line. For the alternative boundary conditions, we choose only temperatures which yield a bound ($\Bernoulli<0$) outer boundary condition, since gas with $\Bernoulli>0$ will likely escape from the halo rather than form an accretion flow. 
As can be seen in the plot, the solutions deviate from the marginally-bound solution only near the outer boundary, with a maximum deviation in temperature at $100\kpc$ of $\pm0.15\,{\rm dex}$. A similar conclusion is reached if we impose some gas density at the outer boundary. The exact choice of the outer boundary condition thus does not affect our conclusions. 

\begin{figure*}
 \includegraphics{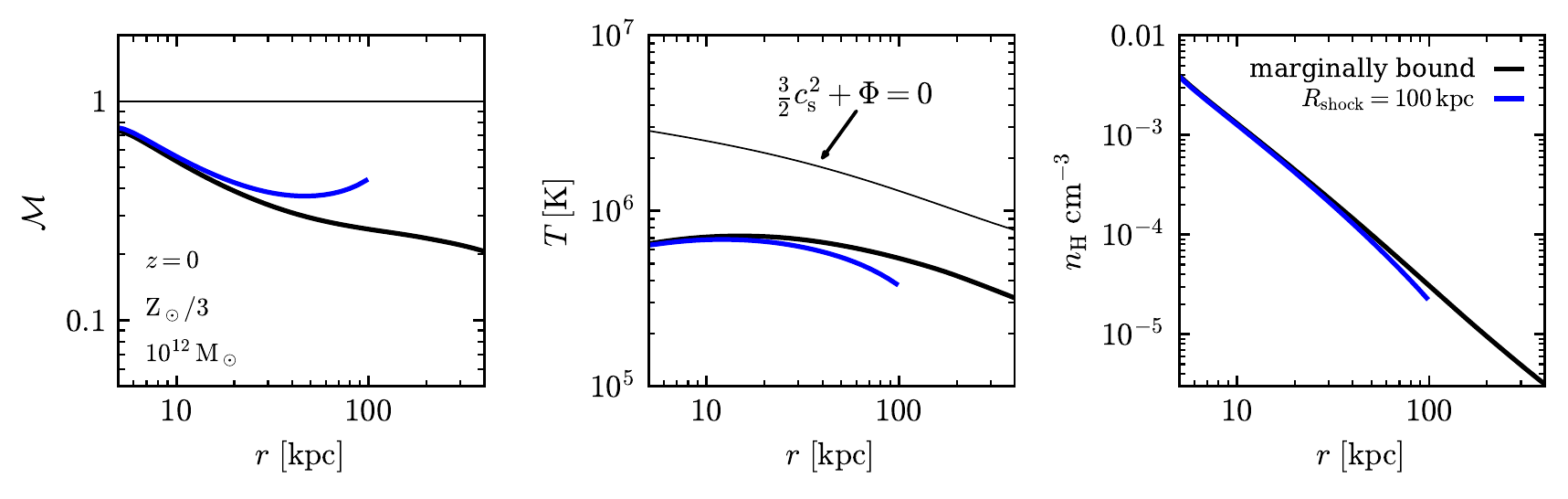}
\caption{Comparison of a marginally-bound solution with a solution which satisfies shock jump conditions at $R_{\rm shock}=100\kpc$. The solutions have the same $\Mdot$. The solutions differ mainly at $\gtrsim 100\kpc$.}
\label{f:t_ratio_shock}
\end{figure*}

How would the solution change if the outer boundary is an accretion shock?
In Figure~\ref{f:t_ratio_shock} we compare a marginally-bound solution with the  solution discussed in section~\ref{s:discrepancy}, which satisfies shock jump conditions at $R_{\rm shock}=100\kpc$. 
Both solutions have the same $\Mdot$. The solutions differ mainly at $\lesssim100\kpc$, consistent with the conclusion from Fig.~\ref{f:outer boundary}.

\section{The timescale for establishing the $\tcool\approx\tflow$ relation}\label{a:t_ratio}
Figures~\ref{f:M15 sims} -- \ref{f:M12 sims high density} show that in the hydrodynamic simulations $\tcool\approx\tflow$ out to radii $r\gg\Rcool$, indicating that this relation is established on a timescale $\ll \tcool$. In this section we explain the origin of this behavior. 

At times $\tsim \ll \tcool$, the radial momentum equation of an initially hydrostatic halo is 
\begin{equation}
 \frac{\d v}{\d t} = -\frac{1}{\rho}\frac{\partial P}{\partial r} - \frac{\vc^2}{r} \approx -\frac{1}{\rho}\frac{\partial (P_0 (1-\tsim/\tcool))}{\partial r} - \frac{\vc^2}{r}
\end{equation}
where $P_0$ is the gas pressure at $t=0$ which satisfies $\rho^{-1}\partial P_0/\partial r = \vc^2 / r$.  The flow acceleration hence equals
\begin{equation}
 \frac{\d v}{\d t} = \frac{P_0\tsim}{\rho}\frac{\partial (\tcool^{-1})}{\partial r} \approx -\frac{r^2}{r^2}\frac{\cs^2}{\gamma}\frac{t}{\tcool^2}\frac{\partial\tcool}{\partial r} \sim -\frac{r/\tcool}{\tff}\frac{t}{\tff}\frac{\partial\ln\tcool}{\partial\ln r}
\end{equation}
where in the last `$\sim$' we used $r/\cs\sim\tff$ and disregarded order-unity factors. At $t\approx\tff$, the term on the right is of order $r/(\tcool\tff)$, indicating that a velocity of order $r/\tcool$ can be reached within a free-fall timescale. The $\tcool\approx\tflow$ relation is thus established on a sound-crossing timescale rather than on a cooling timescale.

\bsp	
\label{lastpage}

\end{document}